\date{}
\documentclass[aps,pra,showpacs]{revtex4-1}
\usepackage{graphicx,amsmath,amssymb,amsfonts,latexsym,color,dcolumn}
\usepackage{soul}

\begin{document}

\title{Landauer's formula breakdown for radiative heat transfer and non-equilibrium Casimir forces}

\author{Adri\'an E. Rubio L\'opez$^{1,2}$\footnote{arubio@df.uba.ar}, Pablo M. Poggi$^3$, Fernando C. Lombardo$^3$, Vincenzo Giannini$^{4,5}$}

\affiliation{$^1$Institute for Quantum Optics and Quantum Information of the Austrian Academy of Sciences, Technikerstrasse 21a, Innsbruck 6020, Austria\\$^2$Institute for Theoretical Physics, University of Innsbruck, A-6020 Innsbruck, Austria\\$^3$Departamento de F\'\i sica Juan Jos\'e
Giambiagi, FCEyN UBA and IFIBA CONICET-UBA, Facultad de Ciencias Exactas y Naturales,
Ciudad Universitaria, Pabell\' on I, 1428 Buenos Aires, Argentina\\$^4$Department of Physics, Condensed Matter Theory, Imperial College London, London SW7 2AZ, United Kingdom\\$^5$Instituto de Estructura de la Materia, IEM-CSIC, E-28006 Madrid, Spain}

\date{today}

\begin{abstract}
In this work we analyze the incidence of the plates' thickness on the Casimir force and radiative 
heat transfer for a configuration of parallel plates in a non-equilibrium scenario, relating to 
Lifshitz's and Landauer's formulas. From a first-principles canonical quantization scheme for the study of
the matter-field interaction, we give closed-form expressions for the non-equilibrium Casimir force and 
the heat transfer between plates of thickness $d_{\rm L},d_{\rm R}$. We distinguish three different contributions to the Casimir force and to the heat transfer 
in the general non-equilibrium situation: two associated to each of the plates, and one to the initial state of the field. We analyze the dependence of the Casimir force and heat transfer with the 
plate thickness (setting $d_{\rm L}=d_{\rm R}\equiv d$), showing the scale at which each magnitude 
converges to the value of infinite thickness ($d\rightarrow+\infty$) and how to correctly reproduce 
the non-equilibrium Lifshitz's formula. For the heat transfer, we show that Landauer's formula does 
not apply to every case (where the three contributions are present), but it is correct for some 
specific situations. We also analyze the interplay of the different contributions for realistic 
experimental and nanotechnological conditions, showing the impact of the thickness in the 
measurements. For small thickness (compared to the separation distance), the plates act to decrease 
the background blackbody flux, while for large thickness the heat is given by the baths' 
contribution only. The combination of these behaviors allows for the possibility, on one hand, of having 
a tunable minimum in the heat transfer that is experimentally attainable and observable for metals, 
and, on the other hand, of having vanishing heat flux in the gap when those difference are of 
opposite signs (thermal shielding). These features turns out to be relevant for nanotechnological 
applications.
\end{abstract}

%\pacs{03.70.+k; 03.65.Yz; 42.50.-p}

\maketitle

%%%%%%%%%%%%%%%%%%%%
\section{Introduction}
%%%%%%%%%%%%%%%%%%%%

One of the most fundamental aspects of every physical theory describing quantum phenomena is dispersion, which unavoidably emerges in every formalism considered. This means that, at the quantum level, we always deal with dynamics that include background fluctutations, which enforces the employment of statistical quantities to describe the reality of nature.

Within this context, fluctuation features are found for every state of the system 
under study, even for the state of lowest energy. This state is usually called `ground state' although 
in contexts involving the notion of particles it is also referred to as `vacuum state', since it 
commonly corresponds to the state with zero number of particles or, in other words, without real 
particles. However, the statistical aspects allow for the concept of virtual particles, characterized by 
a ephemeral existence and then being part of fluctuational deviations of the particle number with 
respect to the number of real particles. Nevertheless, although these virtual particles seem to 
have no physical reality, the fluctuational deviations are precisely responsible of purely quantum 
phenomena that can be experimentally measured and present no classical equivalent. That is how 
the quantum nature of vacuum takes part in the description of different physical situations (see 
Refs.\cite{Milonni,BreuerPetruccione}).

Within the context of quantum field theory (QFT), dispersion phenomena includes van der 
Waals-Casimir forces and heat transfer between micro- and macroscopic bodies, including micro and 
nano-electromechanical systems (MEMS and NEMS, see Ref.\cite{Capasso}). Thus, the possibility of 
having completely quantum effects appearing at macroscales is a reality and studying them it is of 
relevant interest from the theoretical, experimental and technological points of view (see 
Ref.\cite{BordagMohideenMostepanenko}).

In the case of the van der Waals-Casimir force, there is a vast number of remarkable works 
analyzing different aspects related to multiple configurations, geometries (see for example 
Refs.\cite{Levin,Rodriguez, gies,cyl1,cyl2,de}) and materials in different contexts (see Refs.\cite{KardarGolestanian,Lamoreaux,VolokitinPersson,KlimchiMohMoste,dalvitdielectrico} to mention a few). 
Moreover, there are also works studying thermodynamic aspects of the Casimir force involving dissipative materials (see Refs.\cite{IntraLam,PiroLam,LuoZhaoPendry,BimonteEmigKardarKruger,Bordag2017}).

The previous research includes equilibrium and non-equilibrium situations. It should be noted that in the context of the dispersion phenomena addressed in this work (static Casimir forces and steady heat flux), `non-equilibrium' stands for steady situations (with time-independent quantities) that cannot be described as a thermodynamic equilibrium scenario between the parts of the total system considered.

In the Casimir framework, it must be mentioned first the pioneer work of Lifshitz (see Ref.\cite{Lifshitz1956}) that was the first to shed light on the theoretical framework which allows to include dissipative materials in the calculation of Casimir forces. In that work he developed the basis of the currently called Fluctuational Quantum Electrodynamics (FQED) by combining the electrodynamics in real media with the quantum fluctuation-dissipation theorem applied to the current sources at zero temperature. This demonstrate the fact that the Casimir force, existing even at zero temperature, is a macroscopic effect of quantum origin. The configuration that he analyzed was conformed by two parallel plates of infinite thickness (or half-spaces) separated by a vacuum gap. Although his result was valid for zero temperature, the finite-temperature generalization for half-spaces did not take long to be achieved also by Lifshitz and other author, but this time from a fully QFT approach (see Ref.\cite{DzyLifPita}). The expression found for the force gives what it is now commonly referred as `Lifshitz formula'. In that work the expression was given in the Matsubara representation. Nevertheless, the formula can be also written as an integral whose integrand is a product of two factors, one containing the information of the materials in the reflection coefficients of each half-space and another one including the temperature as the only parameter. Moreover, it was shown that in thermal equilibrium the expression for the Casimir force between finite-width plates results with the same form of the Lifshitz formula but the reflection coefficients are the ones for plates of finite thickness. Throughout the work we will use `thickness' and `width' indistinctly for referring to the length of the parallel plates in the normal direction to its surfaces. Clearly, Lifshitz's formula is re-obtained from this result by taking the infinite-thickness limit for the plates, which is guaranteed by the fact that the finite-width reflection coefficients reduce to the ones for half-spaces (see Ref.\cite{KlimchiMohMoste} for a review on this).

It should be noted that, throughout Casimir physics, a crucial point is always to handle and substract infinites that arise unavoidably in QFTs. In other words, for obtaining finite results, a regularization procedure has to be implemented. There are different methods for handling divergences depending on the situation analyzed. Beyond them, for parallel plates made of dissipative materials, no infrared divergences occur and the ultraviolet divergences are prevented by the natural cut-off provided by the dissipation in the material (see for example Ref.\cite{LomMazzRL}), which takes in account the fact that the materials are transparent for high frequencies (for the case without dissipation, a cut-off function has to be introduced by hand to obtain a finite result). However, the result is infinite due to the inherent zero-point fluctuations. These divergences are contained in the mentioned factor associated to the materials in the integral form for the force. For two finite-width plates in thermal equilibrium, there are two methods for eliminating these divergences. One is the `Casimir prescription' (see Ref.\cite{Milonni}), consisting in calculating the energy contained in the gap between the plates and substracting it with the energy contained in the same region and in the same situation (thermal equilibrium) for the case of free-space, i.e., in absence of the plates. This method also applies for half-spaces. The other method is based on substracting the radiation pressure at each side of one of the (finite-width) plates, which corresponds to the net force over the given plate (see Ref.\cite{RubioLopez2017} for example). Both methods gives the mentioned finite-width formula at thermal equilibrium. Moreover, it is worth noting that both regularization procedures takes into account the specific scenario of thermal equilibrium in order to only affect the factor associated to the materials in a correct way.

On the other hand, for non-equilibrium scenarios, one can found research addressing configurations involving point-dipoles, spheres and half-spaces (see Refs.\cite{BimonteEmigKardarKruger,Antezza,BehuHu,IntraBehu,KrugerBimonteEmigKardar,BenJou,BenBiehsJou} and literature cited therein) based fundamentally in FQED approach. A first full QFT approach to non-equilibrium scenarios was recently developed in Ref.\cite{CTPGauge}, and then successfully implemented to derive the half-spaces' result obtained from FQED (see Ref.\cite{TuMaRuLo}). Also in these situations, regularization procedures are required. In FQED formalisms, this step is typically accomplished by discarding the bulk part of the Green tensor, which ensures that all the terms independent of the relative positions of the material bodies are effectively discarded. Thus, this can be seen as a third method that apply for very general situations, but within the context of a FQED approach (see also Refs.\cite{MessAnt,Buhmann}).

To the best of our knowledge there are no previous works dedicated to investigate whether the two methods described above for the study of finite-width plates in equilibrium can be implemented in situations out of equilibrium and how to do it in a conceptually clear approach. Moreover, the non-equilibrium version of the Lifshitz formula was obtained but not deduced from the finite-width case within the context of a full QFT approach. One recent work going in this direction is Ref.\cite{RubioLopez2017}, where a canonical quantization formalism is developed to obtain the force between finite-width plates in a non-equilibrium scenario characterized by thermal and squeezed states, but not addressing the previous question. Within this framework, the Casimir force is given by two types of contributions, one associated to the radiation generated by the plates and another one associated with the initial state of the field.

Now, we consider this approach to give a clear answer to this issue as part of the results of the present work. By considering an initial state for the field that gives a temperature for modes impinging the plates configuration from the left and another one to the modes impinging from the right, we show how in a non-equilibrium scenario the connection between the finite-width result and the non-equilibrium Lifshitz formula is achieved. This shows how to adapt the Casimir prescription for these situations, while the method based on the pressures subtraction leads to an incorrect result.

On the other hand, for the case of heat transfer, the phenomenon admits classical or quantum descriptions depending on the particular scenario. In the far-field, there are plenty of works describing the heat transfer in terms of classical frameworks based on electrodynamics combined with thermodynamics results (see Ref.\cite{HeatHowell}). The heat transfer between bodies at different temperatures is given by Stefan's law of heat exchange (see also Refs.\cite{LandsbergDeVos,PerRubLap}), which is basically the difference between the blackbody radiation emitted by each body, only considering the statistical properties of propagating modes of the electromagnetic (EM) field. Additionally, when the macroscopic bodies have a particular shape, one can develop a theory implementing this law but weighted with the geometrical properties of the given configuration. This approach basically takes into account how much radiation emitted from one body impinges the other. Within this framework, the radiation is treated in a thermodynamical way, without giving a specific description as EM waves.

As the typical distances involved in the situation addressed get shorter, far-field treatment is no longer valid and the wave nature of the EM field begins to become crucial. Propagating modes lead the heat exchange, but diffraction and interference-like phenomena could have an impact on some configurations. Moreover, in the near-field regime, the evanescent modes start to contribute becoming a channel of heat transfer that cannot be neglected and, in fact, could be the most important contribution in some situations. Then, a wave description of the EM radiation is mandatory. Moreover, when considering dissipative materials from first-principles models, low temperatures or entirely quantum objects (as magnetic moments of spin), the need of a quantum theory becomes relevant. Some aspects can be described in a semiclassical way, through a stochastic electrodynamic theory, but replacing the classical fluctuation-dissipation theorem by its quantum version, as it happens for FQED (see Refs.\cite{Milonni,Buhmann}). However, the development of a purely quantum approach enriches from a conceptual point of view and allows the study of regimes beyond the classical (see for example Refs.\cite{Philbin,Barton2016,Manjavacas2017}).

The analog to Lifshitz's work for Casimir forces but in heat transfer is Ref.\cite{PolderVanHove}. There, the authors developed a general EM theory and deduced the heat exchange between two half-spaces at different temperature. The approach includes propagating and evanescent modes' contributions and include a quantum fluctuation-dissipation relation for the sources of current. The result obtained for the heat transfer has the form of Landauer's formula, where the heat is expressed as an integral over the frequencies with its integrand given by a product of two factors, one given by the difference of the boson occupation numbers of the radiations emitted by each body, and another one including the geometrical and material properties of the bodies. This formula also predicts the enhancement of heat transfer in the near-field regime due to the growing of the evanescent modes' contribution. However, although is widely used for different scenarios and configurations (see also Refs.\cite{BenAbdallah1,BenAbdallah2}), another of the main achievements of the present work is to show that this formula is not valid in general for a finite-width plate configuration. Moreover, we show in which cases a Landauer's formula is obtained, gaining intuition about the physical properties of the different contributions that appear. Within the same scenario considered for the calculations regarding the force, we show that for the heat flux between the plates, Landauer's formula is not obtained for finite thickness, even if the initial state of the field is taken as the vacuum state (zero temperature). This is conceptually different from what it is analyzed in Refs.\cite{BenAbdallah1,BenAbdallah2}. We consider that this understanding is crucial for the correct design of experiments at the micro- and nanoscale and also for the development and improvements of novel nanotechnological devices as MEMS and NEMS and, moreover, involving typical metals (see Ref.\cite{Lamoreaux}).

All these features are studied and complemented with numerical calculations, exploiting some interesting phenomena that are expected with the physical intuition obtained.

In order to focus the main text of this work and the calculations on the mentioned results and the 
numerical calculations, we have taken the formal results of Ref.\cite{RubioLopez2017} as a starting 
point and left some specific calculations and deductions to appendixes at the end of the paper. The 
paper is organized as follows: in the next section we summarize the model, the field equation and 
the steady solution for the field operator obtained in Ref.\cite{RubioLopez2017}. In Sec. III, we 
summarize the separation in contributions of the expectation value of the energy-momentum tensor, 
calculating general forms for arbitrary bodies but assuming thermal states for each bath and 
introducing an intrinsic non-equilibrium initial state for the field (the properties of this 
particular initial state for the field are also described in Appendix A). Then, we use these 
features to obtain the Casimir force and the heat flux between two plates of different thickness and 
materials (in Appendix B there are formulas complementing the obtained result). In Sec. IV, we study 
different aspects of the general formulas, showing that they reproduce all the previous (and 
well-known) results as particular cases, including Lifshitz's and Landauer's formulas. In Appendix 
C, there are some complementary calculations to this section. Section V is devoted to the numerical 
analysis for the case of identical plates (same material and thickness), showing the scale of 
convergence to the infinite-thickness expressions (given by Lifshitz's and Landauer's formulas), the 
possibility of tuning the heat flux between the plates even to zero. Finally, Sec. VI summarizes 
our findings.

For simplicity, we have set $\hbar = k_{B} = c = 1$.

%%%%%%%%%%%%%%%%%%%%
\section{Model, Time-Evolution and Steady State}
%%%%%%%%%%%%%%%%%%%%

One way to address the matter-field interaction at a quantum level is to give a first-principles microscopic model for describing the quantum field in interaction with quantum degrees of freedom at each point of space (representing matter). In order to include effects of dissipation and noise in the description, we will use the theory of open quantum systems, and treat the full field dynamics having in mind the paradigmatic example of the quantum Brownian motion (QBM) \cite{BreuerPetruccione}.

\begin{figure}
\centering
\includegraphics[width=0.75\linewidth]{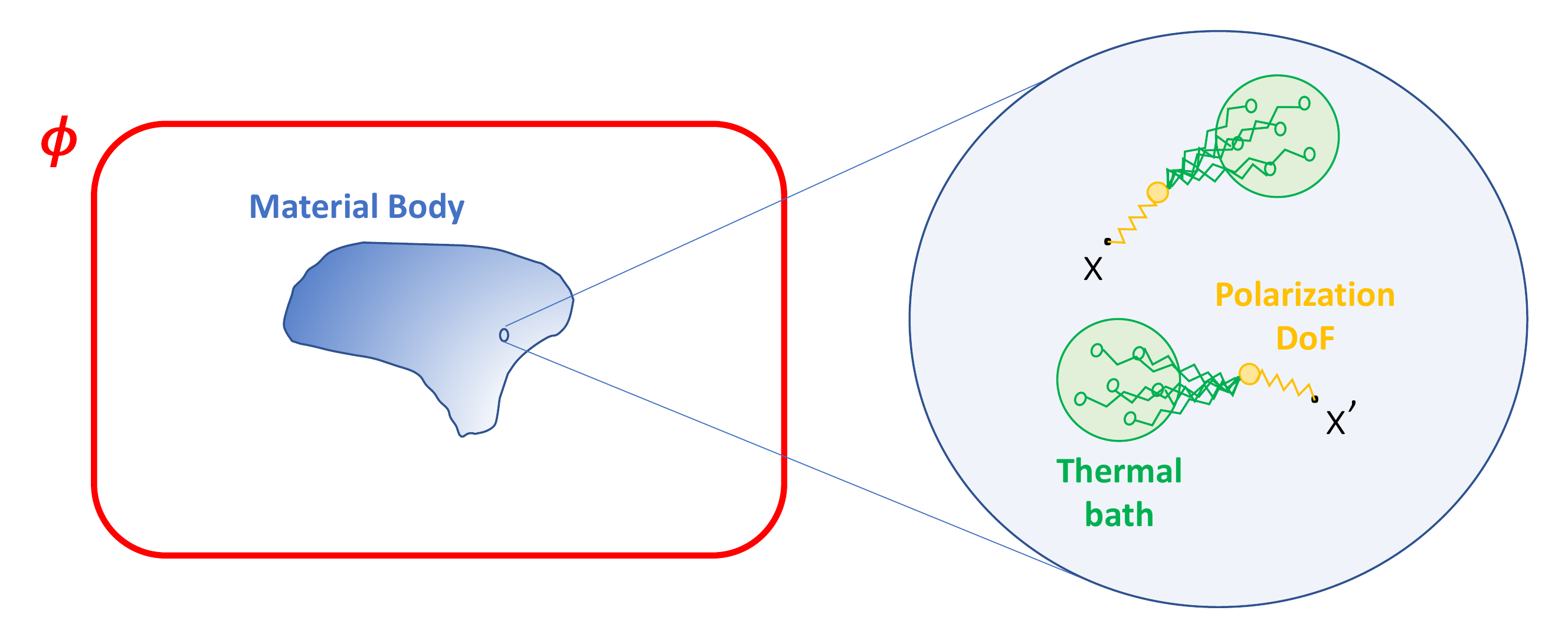}
\caption{Scheme of the composite system considered for the interaction field-matter, expressed in the different terms of the Lagrangian in Eq.(\ref{LagrangianaTOTAL}). It consists in two parts: the field $\phi$ and the material body. At the same time, each point of the material body is given by a polarization degree of freedom coupled to its own thermal bath (a set of harmonic oscillators).}
\label{Esquema}
\end{figure}

In the present work, we consider the same model as the one employed in Ref.\cite{RubioLopez2017}, 
that consists in a system composed of two parts: a massless scalar field and a dielectric material 
which, in turn, are described by their internal degrees of freedom (a set of harmonic oscillators), see Fig.\ref{Esquema}. 
Both sub-systems conform a composite system which, in each point of space, is coupled to a second 
set of harmonic oscillators, that plays the role of an external environment or thermal bath. For 
simplicity we will work in $1+1$ dimensions. In our toy model the massless field represents the 
electromagnetic field, and the first set of harmonic oscillators directly coupled to the scalar 
field represents the polarizable volume elements of the material. In this model the field and the 
volume elements of the material couples through a current-type one (mimicking the typical 
interaction term between the electromagnetic field and matter). The coupling constant for this 
interaction is the electric charge $e$. We will also assume that there is no direct coupling between 
the field and the thermal bath. Thus, the Lagrangian density is considered as

\begin{eqnarray}
\cal{L}&=&\cal{L}_{\phi}+\cal{L}_{S}+\cal{L}_{\phi-S}+\cal{L}_{B}+\cal{L}_{S-B}\nonumber\\&=&\frac{1}{2}~
\partial_{\mu}\phi~\partial^{\mu}\phi+4\pi\eta\left(\frac{1}{2}~m~\dot{r}^{2}_{x}(t)-\frac{1}{2}~m\omega_{0}^{2}~r^{2}_{x}(t)\right)+4\pi\eta
e~\phi(x,t)~\dot{r}_{x}(t)\nonumber\\&+&4\pi\eta\sum_{n}\left(\frac{1}{2}~m_{n}~\dot{q}_{n,x}^{2}(t)-\frac{1}{2}~m_{n}\omega_{n}^{2}~q_{n,x}^{2}(t)\right)-4\pi\eta\sum_{n}\lambda_{n}~q_{n,x}(t)~r_{x}(t),
\label{LagrangianaTOTAL}
\end{eqnarray}

\noindent where the different terms on the r.h.s. correspond to the different parts of the total 
composite system and its interactions. The first term corresponds to the Lagrangian of the massless 
scalar field. The second one, containing brackets, accounts for the polarization degrees of freedom 
of each volume element of the material, described as harmonic oscillators. The third contribution is 
the current-type interaction between the field and the degrees of freedom of the material. The 
fourth, also containing brackets, corresponds to the set of harmonic oscillators conforming the 
thermal bath. The last term is the (linear) interaction of the bath's oscillators with its 
respective volume element degree of freedom.

We have denoted the fact that $r$ and $q_{n}$ have a dependence on position with a label identifying the point of space at which they are located (but it is important to stress that this label is not a dynamical variable; as it happens for the scalar field). It is clear that each atom interacts with a thermal bath placed at the same position. We have denoted by $\eta$ the density of the degrees of freedom of the volume elements. The constants $\lambda_{n}$ are the coupling constants between the volume elements and the bath oscillators. It is implicitly understood that Eq.(\ref{LagrangianaTOTAL}) represents the Lagrangian density inside the material, while outside the Lagrangian is given by the one of a free field.

The quantization of the theory is straightforward. It should be noted that the full Hilbert space ${\cal H}$ of the model is not only the field Hilbert space ${\cal H}_{\phi}$ (as it is considered in others works where the field is the only relevant degree of freedom), but also includes the Hilbert spaces of the volume elements' degrees of freedom ${\cal H}_{\rm A}$ and the bath oscillators ${\cal H}_{\rm B}$, in such a way that ${\cal H}={\cal H}_{\phi}\otimes {\cal H}_{\rm A}\otimes {\cal H}_{\rm B}$. We will assume, as frequently done in the context of QBM,  that for $t<t_{0}$ the three parts of the systems are uncorrelated and not interacting. Interactions are turned on at $t=t_{0}$. Therefore, the initial conditions for the operators $\widehat{\phi}$, $\widehat{r}$ must be given in terms of operators acting in each part of the Hilbert space. The interactions will make that initial operators to  become operators over the whole space ${\cal H}$. The initial density matrix of the total system is of the form:

\begin{equation}
\widehat{\rho}(t_{0})=\widehat{\rho}_{\rm IC}(t_{0})\otimes\widehat{\rho}_{\rm A}(t_{0})\otimes\widehat{\rho}_{\rm B}.
\label{EstadoInicial}
\end{equation}

In principle, each part of the whole system can be in any initial state.
Then, following Ref.\cite{RubioLopez2017} we can straightforwardly write the Heisenberg equations of 
motion  and solve those related to the material's degrees of freedom and introduce it in the 
corresponding field equation to obtain an effective equation for the full dynamics of the field 
operator:

\begin{equation}
\square\widehat{\phi}+\frac{\partial^{2}}{\partial t^{2}}\left[\int_{t_{0}}^{t}d\tau\chi_{x}(t-\tau)\widehat{\phi}(x,\tau)\right]=4\pi\eta
eC(x)\left[\ddot{G}_{2}(t-t_{0})\widehat{r}_{x}(t_{0})+\dot{G}_{2}(t-t_{0})\frac{\widehat{p}_{x}(t_{0})}{m} +\int_{t_{0}}^{t}d\tau~\dot{G}_{2}(t-\tau)\frac{\widehat{F}_{x}(\tau-t_{0})}{m}\right],
\label{EcMovCampoTOTALKnoll}
\end{equation}

\noindent where $\chi_{x}(t)=\omega_{\rm Pl}^{2}~G_{2,x}(t)~C(x)$ is the susceptibility function with $\omega_{\rm Pl}^{2}=\frac{4\pi\eta e^{2}}{m}$ the plasma frecuency and $G_{2,x}$ is the retarded Green function associated to the QBM equation at the point $x$, $\widehat{r}_{x}(t_{0})$ and $\widehat{p}_{x}(t_{0})$ are the position and momentum operator of the volume element degrees of freedom of the material, while $\widehat{F}_{x}$ is the stochastic force operator generated by the bath at $x$ which acts over the corresponding volume element. As it can be seen in Ref.\cite{RubioLopez2017}, this operator is a generalization of the stochastic force operator found in the quantum brownian theory (within an open quantum system framework, see Ref.\cite{BreuerPetruccione}) and it is characterized by its correlations given by a fluctuation-dissipation relation:

\begin{equation}
\Big\langle\left\{\widehat{\overline{F}}^{\infty}_{x'}(\omega'),\widehat{\overline{F}}^{\infty}_{x''}(\omega'')\right\}\Big\rangle_{\rm B}=(2\pi)^{2}~\delta(x'-x'')~\frac{J(\omega')}{2\eta}~\coth\left(\frac{\beta_{{\rm B},x'}\omega'}{2}\right)\delta(\omega'+\omega''),
\end{equation}

\noindent where $\beta_{{\rm B},x'}$ corresponds to the inverse temperature of the thermal bath located at $x'$ (see Fig.\ref{Esquema} and Ref.\cite{RubioLopez2017} for more details).

It is worth noting that we have included an spatial label denoting the straightforward generalization to inhomogeneous media, where each point of the material can have different properties. Beyond this dependence, the boundaries of the material bodies enters through the spatial material distribution function $C$, which is zero in free space points. The regions filled (and the contours) with real material are defined by this function. This is clearly essential for the determination of the field's boundary conditions.

Eq.(\ref{EcMovCampoTOTALKnoll}) can be solved in terms of the retarded Green function 
$\mathfrak{G}_{\rm Ret}$ after initial conditions for the field operator are given. In 
Ref.\cite{RubioLopez2017} this procedure has been done by giving free field initial conditions for 
the field operator, which are expressed in terms of the annihilation and creation operators 
($\widehat{a}_{k}(t_{0}),\widehat{a}_{k}^{\dag}(t_{0})$) of the free field. Therefore, at the 
initial time $t_{0}$, the field is only an operator acting on ${\cal H}_{\phi}$ but the switching-on of the 
interactions causes the field operator to become an operator which acts on the full Hilbert space ${\cal H}$ 
during the time evolution:

\begin{equation}
\widehat{\phi}(x,t)=\widehat{\phi}_{\rm IC}(x,t)\otimes\mathbb{I}_{\rm A}\otimes\mathbb{I}_{\rm B}+\mathbb{I}_{\phi}\otimes\widehat{\phi}_{\rm A}(x,t)\otimes\mathbb{I}_{\rm B}+\mathbb{I}_{\phi}\otimes\mathbb{I}_{\rm A}\otimes\widehat{\phi}_{\rm B}(x,t).
\end{equation}

However, as we are interested in the expressions for the heat transfer and the Casimir force in non-equilibrium but steady situations, we require the long-time limit ($t_{0}\rightarrow-\infty$) of the total field operator. The full expressions for each part during the time evolution and also the deduction of its long-time expressions for the present model can be found in Ref.\cite{RubioLopez2017}, and we obtain:

\begin{equation}
\widehat{\phi}(x,t)\longrightarrow\widehat{\phi}^{\infty}(x,t)=\widehat{\phi}_{\rm IC}^{\infty}(x,t)\otimes\mathbb{I}_{\rm A}\otimes\mathbb{I}_{\rm B}+\mathbb{I}_{\phi}\otimes\widehat{\phi}_{\rm A}^{\infty}\otimes\mathbb{I}_{\rm B}+\mathbb{I}_{\phi}\otimes\mathbb{I}_{\rm A}\otimes\widehat{\phi}_{\rm B}^{\infty}(x,t),
\label{SteadyFieldOperator}
\end{equation}

\noindent with each long-time operator given by:

\begin{eqnarray}
%\widehat{\phi}_{\rm IC}^{(+)}(x,t)\rightarrow
\widehat{\phi}_{\rm IC}^{(+),\infty}(x,t)&=&\int_{-\infty}^{+\infty}dk\left(\frac{1}{\omega_{k}}\right)^{1/2}\widehat{a}_{k}(-\infty)\left[e^{-ikt}~\Theta(k)~\Phi_{-ik}^{>}(x)+e^{ikt}~\Theta(-k)~\left(\Phi_{-ik}^{<}(x)\right)^{*}\right]
%\nonumber\\
%&&+~\left[\text{Time and spatial independent oscillatory term}\right].
,
\label{LongTimeFieldOperatorIC}
\end{eqnarray}

\begin{eqnarray}
%\widehat{\phi}_{\rm A}^{(+)}(x,t)\longrightarrow
\widehat{\phi}_{\rm A}^{(+),\infty}&=&-\frac{1}{2}\int dx'~\frac{4\pi\eta eC(x')}{\sqrt{2m\omega_{0}}}~\widehat{b}_{0,x'}(-\infty),
\label{LongTimeFieldOperatorA}
\end{eqnarray}

\begin{equation}
%\widehat{\phi}_{\rm B}(x,t)\longrightarrow
\widehat{\phi}_{\rm B}^{\infty}(x,t)=\int dx'~\frac{4\pi\eta eC(x')}{m}\int_{-\infty}^{+\infty}\frac{d\omega}{2\pi}~e^{-i\omega t}~i\omega~\overline{G}_{2}(\omega)~\overline{\mathfrak{G}}_{\rm Ret}(x,x',\omega)~\widehat{\overline{F}}^{\infty}_{x'}(\omega),
\label{LongTimeFieldOperatorB}
\end{equation}

\noindent where $\Phi^{\lessgtr}$ are the homogeneous solutions associated to the homogeneous field 
equation and satisfying only the boundary condition on each limit of the variable value's interval, 
$\widehat{b}_{0,x'}(-\infty)$ is the annihilation operator of the volume element degree of freedom, 
while the over-lines are denoting Fourier transforms.

%%%%%%%%%%%%%%%%%%%%
\section{Force and Radiative Heat Exchange at the Steady State}
%%%%%%%%%%%%%%%%%%%%

With the field operator at the steady state, we can evaluate both the Casimir force and the heat transfer between two plates in an unified way by calculating the expectation values of the energy-momentum tensor operator. The quantum version of the energy-momentum tensor is obtained by symmetrizing the classical expression after promoting the field to a quantum operator, giving:

\begin{equation}
\widehat{T}_{\mu\nu}(x_{1}^{\sigma},t_{0})\equiv\left(\delta_{\mu}^{~\gamma}\delta_{\nu}^{~\alpha}-\frac{1}{2}~\eta_{\mu\nu}\eta^{\gamma\alpha}\right)\frac{1}{2}\left(\partial_{\gamma}\widehat{\phi}(x_{1}^{\sigma})~\partial_{\alpha}\widehat{\phi}(x_{1}^{\sigma})+\partial_{\alpha}\widehat{\phi}(x_{1}^{\sigma})~\partial_{\gamma}\widehat{\phi}(x_{1}^{\sigma})\right).
\end{equation}

As the field operator in the steady state is given by Eq.(\ref{SteadyFieldOperator}), by noting that the contribution associated to the volume elements is independent of time and space, we have that for the derivatives holds $\partial_{\mu}\widehat{\phi}^{\infty}=\partial_{\mu}\widehat{\phi}_{\rm IC}^{\infty}\otimes\mathbb{I}_{\rm A}\otimes\mathbb{I}_{\rm B}+\mathbb{I}_{\phi}\otimes\mathbb{I}_{\rm A}\otimes\partial_{\mu}\widehat{\phi}_{\rm B}^{\infty}$. Therefore, the volume elements has no contribution to the expectations values of the energy-momentum tensor. Moreover, as shown in Ref.\cite{RubioLopez2017}, the expectation values of the annihilation and creation operators are zero for thermal states, and we are considering thermal states for the baths. This turns to be enough to prove that the expectation value of the energy-momentum tensor splits into two contributions, one associated to the initial conditions of the field and the other one associated to the baths:

\begin{equation}
\Big\langle\widehat{T}_{\mu\nu}^{\infty}(x_{1}^{\sigma})\Big\rangle=\Big\langle\widehat{T}_{\mu\nu}^{{\rm IC},\infty}(x_{1}^{\sigma})\Big\rangle_{\phi}+\Big\langle\widehat{T}_{\mu\nu}^{{\rm B},\infty}(x_{1}^{\sigma})\Big\rangle_{\rm B},
\label{ExpectationValueEnergyMomentumTensor}
\end{equation}

\noindent where $\langle...\rangle_{\phi, \rm B}={\rm Tr}_{\phi, \rm B}\left(\widehat{\rho}_{\rm IC, B}...\right)$, denoting that each trace is taken in the corresponding part of the total Hilbert space.

Nevertheless, while for the baths we assume thermal states, for the field we will consider an intrinsic non-equilibrium state that takes into account the possibility for the initial free field to be in a state with net radiation going from left to right. Although the configuration is surrounded by free space, it is of phenomenological interest to consider a scenario where the configuration of plates is in contact with a general reservoir (i.e., the plates are inside an oven) with its left and right walls located at $x=-\infty$ and $x=+\infty$ respectively and having each one at different (inverse) temperature $\beta_{\phi,\rm L}$ and $\beta_{\phi,\rm R}$. Initially, before the appearance of the plates, having this situation clearly generates an intrinsic flow of heat from the hottest wall to the coldest one. After the appearance of the plates, during the transient stage, this flow is modified by the presence of the plates (as it happens in Ref.\cite{RubioLopez2017} for the field in an initial thermal state) until reaching the (steady) long-time regime.

Therefore, as the walls of the (hypothetical) oven are held at different temperatures, the crucial point here is that the modes representing traveling-waves from left to right ($k>0$) will radiate at the inverse-temperature $\beta_{\phi,\rm L}$, while the modes representing traveling-waves from right to left ($k<0$) will radiate at the inverse-temperature $\beta_{\phi,\rm R}$. Then, the intrinsic non-equilibrium state for the field will be defined by the expectation values:

\begin{equation}
\Big\langle\widehat{a}_{k}(-\infty)\widehat{a}_{k'}(-\infty)\Big\rangle_{\phi}=0~~~,~~~\left\langle\widehat{a}_{k}^{\dag}(-\infty)\widehat{a}_{k'}(-\infty)\right\rangle_{\phi}=\left[\Theta(k)~N_{\phi,\rm L}(\omega_{k})+\Theta(-k)~N_{\phi,\rm R}(\omega_{k})\right]\delta(k-k'),
\label{AnnihiCreaEVIntrinsicNonEq}
\end{equation}

\noindent where the typical expectation values for a thermal state (see Ref.\cite{RubioLopez2017}) are simply recovered by setting $\beta_{\phi,\rm L}=\beta_{\phi,\rm R}$. More about the intrinsic non-equilibrium state is shown in Appendix \ref{INEISF}. In Fig.\ref{Evolucion}, a scheme of the configuration of plates within our formalism can be found.

\begin{figure}
\centering
\includegraphics[width=1.00\linewidth]{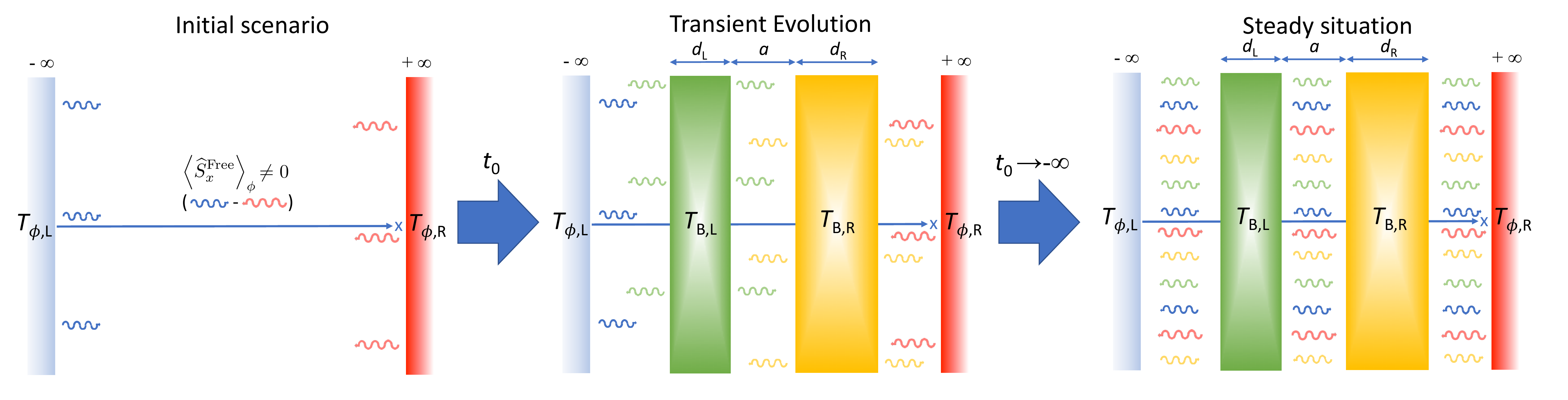}
\caption{Scheme of the configuration of plates within the canonical quantization formalism employed. The initial time corresponds to $t=t_{0}$ and the plates are not interacting with the scalar field. The field is free an it has a state of net heat flux different from zero. This information is encoded in the chosen `intrinsic non-equilibrium state' defined by Eq.(\ref{AnnihiCreaEVIntrinsicNonEq}) and commented in App.\ref{INEISF}. The amount of net heat flux is defined by the temperatures $T_{\phi,\rm L,R}$, that can be interpreted as the temperatures of the walls of the hypothetical oven where the plates are placed. At $t=t_{0}$ the interaction of the field with the plates starts, having a transient stage that reach a steady situation ($t_{0}\rightarrow-\infty$). All the points of each plates are assumed to be at a given temperature $T_{\rm B,L,R}$ (each body has an uniform temperature). In the steady situation, there is radiation emmitted by the plates and by the walls of the hypothetical oven in all the regions and with all directions due to the reflections in the bodies. In this situation the Casimir force and the heat flux are evaluated.}
\label{Evolucion}
\end{figure}

Considering this, for a general configuration, both terms of the expectation value of the components of the energy-momentum tensor can be calculated by employing the Green function and the homogeneous solutions for the given problem, obtaining the following expressions:

\begin{eqnarray}
\Big\langle\widehat{T}_{\mu\nu}^{{\rm IC},\infty}(x_{1}^{\sigma})\Big\rangle_{\phi}&=&\int_{-\infty}^{+\infty}dk\frac{1}{\omega_{k}}\left[\Theta(k)~\coth\left(\frac{\beta_{\phi,\rm L}\omega_{k}}{2}\right)+\Theta(-k)~\coth\left(\frac{\beta_{\phi,\rm R}\omega_{k}}{2}\right)\right]{\rm Re}\Big[\left(\delta_{\mu}^{~0}(-i\omega_{k})\Phi_{k}+\delta_{\mu}^{~1}\Phi_{k}^{'}\right)\nonumber\\
&&\times\left(\delta_{\nu}^{~0}i\omega_{k}\left(\Phi_{k}\right)^{*}+\delta_{\nu}^{~1}(\Phi_{k}^{'})^{*}\right)-\frac{\eta_{\mu\nu}}{2}\Big(\omega_{k}^{2}|\Phi_{k}|^{2}
-|\Phi_{k}^{'}|^{2}\Big)\Big],
\label{TMuNuICThermal}
\end{eqnarray}

\begin{eqnarray}
\Big\langle\widehat{T}_{\mu\nu}^{{\rm B},\infty}(x)\Big\rangle_{\rm B}&=&\int dx'~C(x')\int_{-\infty}^{+\infty}d\omega~\frac{\omega^{2}}{2}~{\rm Re}(n_{x'})~{\rm Im}(n_{x'})\coth\left(\frac{\beta_{{\rm B},x'}\omega}{2}\right)\nonumber\\
&\times&\Big(\left[\delta_{\mu}^{~0}~(-i\omega)+\delta_{\mu}^{~1}~\partial_{x}\right]\overline{\mathfrak{G}}_{\rm Ret}(x,x',\omega)\left[\delta_{\nu}^{~0}~i\omega+\delta_{\nu}^{~1}~\partial_{x}\right]\overline{\mathfrak{G}}_{\rm Ret}^{*}(x,x',\omega)\nonumber\\
&&+\left[\delta_{\mu}^{~0}~(-i\omega)+\delta_{\mu}^{~1}~\partial_{x}\right]\overline{\mathfrak{G}}_{\rm Ret}^{*}(x,x',\omega)\left[\delta_{\nu}^{~0}~i\omega+\delta_{\nu}^{~1}~\partial_{x}\right]\overline{\mathfrak{G}}_{\rm Ret}(x,x',\omega)\nonumber\\
&&-~\eta_{\mu\nu}\left[\omega^{2}|\overline{\mathfrak{G}}_{\rm Ret}(x,x',\omega)|^{2}-|\partial_{x}\overline{\mathfrak{G}}_{\rm Ret}(x,x',\omega)|^{2}\right]\Big),
\label{TMuNuB}
\end{eqnarray}

\noindent where in the initial conditions' contribution we have used the notation that $\Phi_{k}(x)=\Phi_{-ik}^{>}(x)$ for $k>0$ while $\Phi_{k}(x)=(\Phi_{-ik}^{<}(x))^{*}$ for $k<0$, and $\omega_{k}=|k|$.

However, for a specific material configuration the homogeneous solutions $\Phi$ (from which the Green function can be constructed in a straightforward way) have to be calculated. If we consider a configuration of two plates of thickness $d_{\rm L,R}$ respectively  and different homogeneous materials separated by a distance $a$ and surrounded by vacuum, those solutions $\Phi$ can be determined easily (see Ref.\cite{RubioLopez2017}). As we are considering a non-equilibrium situation, the Casimir force will be calculated from the expectation value of the $xx-$component of the energy-momentum tensor, evaluated in the region between the plates and substracting it with the same quantity in absence of the plates configuration. This prescription is exactly the Casimir prescription for regularizing the expression of the force, that here we apply for a non-equilibrium situation. It is worth noting that the method employing the radiation pressures at each sides of one of the plates (as it is done for instance in Ref.\cite{RubioLopez2017} and references therein) is not applicable for this situation since it gives an incorrect regularization for the force and, moreover, different values of the force acting each plate. However, we can say that both approaches agree when the same state (thermal or not) is considered for each plate and for all the modes of the initial conditions' contribution (as it happens in Ref.\cite{RubioLopez2017}). Therefore, the Casimir force is given by:

\begin{equation}
F_{\rm C}=\langle\widehat{T}_{xx}^{\rm Free}\rangle_{\phi}-\langle\widehat{T}_{xx}^{\infty}\rangle^{\rm Int}=\langle\widehat{T}_{xx}^{\rm Free}\rangle_{\phi}-\langle\widehat{T}_{xx}^{{\rm IC},\infty}\rangle^{\rm Int}_{\phi}-\langle\widehat{T}_{xx}^{{\rm B},\infty}\rangle^{\rm Int}_{\rm B},
\label{ForceCasimir}
\end{equation}

\noindent where $\langle\widehat{T}_{xx}^{\rm Free}\rangle_{\phi}$ is given by Eq.(\ref{TxxNonEq}). It is worth mentioning that the temperatures in the regularization term $\langle\widehat{T}_{xx}^{\rm Free}\rangle_{\phi}$ will be taken as $\beta_{\phi,{\rm L,R}}$ since it corresponds to a situation without plates and entirely defined by the walls of the (big) oven.

Therefore, each contribution is given by:

\begin{eqnarray}
\langle\widehat{T}_{xx}^{{\rm IC},\infty}\rangle^{\rm Int}_{\phi}[a,d_{\rm L},d_{\rm R},&\beta_{\phi,\rm L},&\beta_{\phi,\rm R}] \nonumber \\ &=&\int_{0}^{+\infty}dk~k\left[\coth\left(\frac{\beta_{\phi,\rm L}k}{2}\right)\left(\left|C_{-ik}^{>}\right|^{2}+\left|D_{-ik}^{>}\right|^{2}\right)+\coth\left(\frac{\beta_{\phi,\rm R}k}{2}\right)\left(\left|C_{-ik}^{<}\right|^{2}+\left|D_{-ik}^{<}\right|^{2}\right)\right]\nonumber\\
&=&\int_{0}^{+\infty}dk~k\frac{\left[\coth\left(\frac{\beta_{\phi,\rm L}k}{2}\right)|t_{\rm L}|^{2}\left(1+|r_{\rm R}|^{2}\right)+\coth\left(\frac{\beta_{\phi,\rm R}k}{2}\right)|t_{\rm R}|^{2}\left(1+|r_{\rm L}|^{2}\right)\right]}{|1-r_{\rm L}r_{\rm R}~e^{i2ka}|^{2}},
\label{CasimirForceICFull}
\end{eqnarray}

\begin{eqnarray}
&&\langle\widehat{T}_{xx}^{{\rm B},\infty}\rangle^{\rm Int}_{\rm B}\left[a,d_{\rm L},d_{\rm R},\beta_{\rm B,L},\beta_{\rm B,R}\right]= \nonumber \\
&=&\int_{0}^{+\infty}d\omega~\frac{\omega}{2}\Bigg[\coth\left[\frac{\beta_{\rm B,L}\omega}{2}\right]{\rm Re}(n_{\rm L})\frac{(1+|r_{\rm R}|^{2})}{|t_{\rm R}|^{2}}\Big(|E_{-i\omega}^{<}|^{2}e^{-\omega{\rm Im}(n_{\rm L})a}\left[1-e^{-2\omega{\rm Im}(n_{\rm L})d_{\rm L}}\right]\nonumber\\
&&+~|F_{-i\omega}^{<}|^{2}e^{\omega{\rm Im}(n_{\rm L})a}\left[e^{2\omega{\rm Im}(n_{\rm L})d_{\rm L}}-1\right]+2~\frac{{\rm Im}(n_{\rm L})}{{\rm Re}(n_{\rm L})}~{\rm Im}\Big[E_{-i\omega}^{<*}F_{-i\omega}^{<}e^{-i\omega{\rm Re}(n_{\rm L})a}\left(1-e^{-i2\omega{\rm Re}(n_{\rm L})d_{\rm L}}\right)\Big]\Big)\nonumber\\
&&+\coth\left[\frac{\beta_{\rm B,R}\omega}{2}\right]{\rm Re}(n_{\rm R})\frac{(1+|r_{\rm L}|^{2})}{|t_{\rm L}|^{2}}\Big(|E_{-i\omega}^{>}|^{2}e^{-\omega{\rm Im}(n_{\rm R})a}\left[1-e^{-2\omega{\rm Im}(n_{\rm R})d_{\rm R}}\right]+|F_{-i\omega}^{>}|^{2}e^{\omega{\rm Im}(n_{\rm R})a}\left[e^{2\omega{\rm Im}(n_{\rm R})d_{\rm R}}-1\right]\nonumber\\
&&+~2~\frac{{\rm Im}(n_{\rm R})}{{\rm Re}(n_{\rm R})}~{\rm Im}\Big[E_{-i\omega}^{>*}F_{-i\omega}^{>}e^{-i\omega{\rm Re}(n_{\rm R})a}\left(1-e^{-i2\omega{\rm Re}(n_{\rm R})d_{\rm R}}\right)\Big]\Big)\Bigg],
\label{CasimirForceBathFull}
\end{eqnarray}

\noindent and the coefficients for the plates configuration $C^{\lessgtr}_{-ik},D^{\lessgtr}_{-ik},E^{\lessgtr}_{-ik},F^{\lessgtr}_{-ik}$, for each mirror $r_{\rm L,R},t_{\rm L,R}$ and for an interfase $r_{n_{\rm L,R}}$ can be found in the Appendix \ref{Coeff}.

It is worth noting that each contribution results symmetric under the interchange of the subscripts $\rm L$ and $\rm R$, which means that the force has the same absolute value for both plates (with opposite signs on each one) and also that the inverted configuration of plates and oven's walls provides the same forces.

In analogy, the heat between the plates is calculated as the expectation value of the Poynting vector in the region between the plates. In 1+1 dimensions, the Poynting vector has only one component corresponding to minus the $x0-$component of the energy-momentum tensor. Then, the heat presents the same structure of contributions as the Casimir force:

\begin{equation}
Q_{\infty}\equiv\Big\langle\widehat{S}_{x}^{\infty}\Big\rangle=-\Big\langle\widehat{T}_{x0}^{\infty}\Big\rangle=Q^{\rm IC}_{\infty}(a,d_{\rm L},d_{\rm R},\beta_{\phi,\rm L},\beta_{\phi,\rm R})+Q_{\infty}^{\rm B}(a,d_{\rm L},d_{\rm R},\beta_{\rm B,L},\beta_{\rm B,R}),
\end{equation}

\noindent where each contribution is given by:

\begin{eqnarray}
Q^{\rm IC}_{\infty}(a,d_{\rm L},d_{\rm R},\beta_{\phi,\rm L},\beta_{\phi,\rm R})&=&\int_{0}^{+\infty}dk~k\left[\coth\left(\frac{\beta_{\phi,\rm L}k}{2}\right)\left(\left|C_{-ik}^{>}\right|^{2}-\left|D_{-ik}^{>}\right|^{2}\right)-\coth\left(\frac{\beta_{\phi,\rm R}k}{2}\right)\left(\left|C_{-ik}^{<}\right|^{2}-\left|D_{-ik}^{<}\right|^{2}\right)\right]\nonumber\\
&=&\int_{0}^{+\infty}dk~k\frac{\left[\coth\left(\frac{\beta_{\phi,\rm L}k}{2}\right)|t_{\rm L}|^{2}\left(1-|r_{\rm R}|^{2}\right)-\coth\left(\frac{\beta_{\phi,\rm R}k}{2}\right)|t_{\rm R}|^{2}\left(1-|r_{\rm L}|^{2}\right)\right]}{|1-r_{\rm L}r_{\rm R}~e^{i2ka}|^{2}},
\label{QICFull}
\end{eqnarray}

\begin{eqnarray}
&&Q_{\infty}^{\rm B}(a,d_{\rm L},d_{\rm R},\beta_{\rm B,L},\beta_{\rm B,R})=\int_{0}^{+\infty}d\omega~\frac{\omega}{8}\Bigg[\coth\left[\frac{\beta_{\rm B,L}\omega}{2}\right]{\rm Re}(n_{\rm L})\frac{(1-|r_{\rm R}|^{2})}{|t_{\rm R}|^{2}}\Big(|E_{-i\omega}^{<}|^{2}e^{-\omega{\rm Im}(n_{\rm L})a}\left[1-e^{-2\omega{\rm Im}(n_{\rm L})d_{\rm L}}\right]\nonumber\\
&&+~|F_{-i\omega}^{<}|^{2}e^{\omega{\rm Im}(n_{\rm L})a}\left[e^{2\omega{\rm Im}(n_{\rm L})d_{\rm L}}-1\right]+2~\frac{{\rm Im}(n_{\rm L})}{{\rm Re}(n_{\rm L})}~{\rm Im}\Big[E_{-i\omega}^{<*}F_{-i\omega}^{<}e^{-i\omega{\rm Re}(n_{\rm L})a}\left(1-e^{-i2\omega{\rm Re}(n_{\rm L})d_{\rm L}}\right)\Big]\Big)\nonumber\\
&&-\coth\left[\frac{\beta_{\rm B,R}\omega}{2}\right]{\rm Re}(n_{\rm R})\frac{(1-|r_{\rm L}|^{2})}{|t_{\rm L}|^{2}}\Big(|E_{-i\omega}^{>}|^{2}e^{-\omega{\rm Im}(n_{\rm R})a}\left[1-e^{-2\omega{\rm Im}(n_{\rm R})d_{\rm R}}\right]+|F_{-i\omega}^{>}|^{2}e^{\omega{\rm Im}(n_{\rm R})a}\left[e^{2\omega{\rm Im}(n_{\rm R})d_{\rm R}}-1\right]\nonumber\\
&&+~2~\frac{{\rm Im}(n_{\rm R})}{{\rm Re}(n_{\rm R})}~{\rm Im}\Big[E_{-i\omega}^{>*}F_{-i\omega}^{>}e^{-i\omega{\rm Re}(n_{\rm R})a}\left(1-e^{-i2\omega{\rm Re}(n_{\rm R})d_{\rm R}}\right)\Big]\Big)\Bigg].
\label{QBathFull}
\end{eqnarray}

In this case (contrary to what happens to the force), the interchange of L and R is antisymmetric in any of the contributions, denoting the fact that if the configuration of plates and oven's walls is reversed, the flux of heat goes in the opposite direction as it is expected.

Eqs.(\ref{ForceCasimir})-(\ref{QBathFull}) are the main results that we will analyze in the relevant (limit)-cases in order to study which is the 
effect of thickness in the total expressions for both, the force and heat in the non-equilibrium scenario.

Nevertheless, we get a simpler expression for the total heat by employing a relation between parts of the integrands regarding the materials that are  based on the fact that in equilibrium ($\beta_{\phi, \rm L}=\beta_{\phi, \rm R}=\beta_{\rm B,L}=\beta_{\rm B,R}=\beta$) the total heat transfer is zero. In other words, as we have, 

\begin{equation}
Q_{\infty}(a,d_{\rm L},d_{\rm R},\beta,\beta,\beta,\beta)=Q^{\rm IC}_{\infty}(a,d_{\rm L},d_{\rm R},\beta,\beta)+Q_{\infty}^{\rm B}(a,d_{\rm L},d_{\rm R},\beta,\beta)\equiv 0,
\label{TotalHeatFluxEq0}
\end{equation}

\noindent this gives us a relation between the part of the integrands in Eqs.(\ref{QICFull}) and (\ref{QBathFull}) involving the material properties since the thermal factors are the same for every term. Then, using this relation we can write the total heat (in general) by mixing the contributions, 

\begin{eqnarray}
&&Q_{\infty}(a,d_{\rm L},d_{\rm R},\beta_{\phi,\rm L},\beta_{\phi,\rm R},\beta_{\rm B,L},\beta_{\rm B,R})
=\nonumber\\
&&=\int_{0}^{+\infty}dk~2k~\frac{\left[\left[N_{\phi,\rm L}(k)-N_{\rm B,R}(k)\right]|t_{\rm L}|^{2}\left(1-|r_{\rm R}|^{2}\right)-\left[N_{\phi,\rm R}(k)-N_{\rm B,R}(k)\right]|t_{\rm R}|^{2}\left(1-|r_{\rm L}|^{2}\right)\right]}{|1-r_{\rm L}r_{\rm R}~e^{i2ka}|^{2}}\label{QTotalLandauer}\\
&&+\int_{0}^{+\infty}dk~\frac{k}{2}\frac{(1-|r_{\rm R}|^{2})}{|1-r_{\rm L}r_{\rm R}~e^{i2ka}|^{2}}\left[N_{\rm B,L}(k)-N_{\rm B,R}(k)\right]\frac{(1-|r_{n_{\rm L}}|^{2})}{|1-r_{n_{\rm L}}^{2}~e^{i2kn_{\rm L}d_{\rm L}}|^{2}}\Bigg[(1+|r_{n_{\rm L}}|^{2}e^{-2k{\rm Im}(n_{\rm L})d_{\rm L}})(1-e^{-2k{\rm Im}(n_{\rm L})d_{\rm L}})\nonumber\\
&&+\frac{4{\rm Im}(n_{\rm L})~e^{-2k{\rm Im}(n_{\rm L})d_{\rm L}}}{|n_{\rm L}+1|^{2}(1-|r_{n_{\rm L}}|^{2})}{\rm Im}\left[r_{n_{\rm L}}\left(1-e^{i2k{\rm Re}(n_{\rm L})d_{\rm L}}\right)\right]\Bigg],\nonumber
\end{eqnarray}

\noindent  where we have used the fact that the factor containing the temperatures reads $\coth\left[\frac{\beta_{j,\rm L}\omega}{2}\right]-\coth\left[\frac{\beta_{j,\rm R}\omega}{2}\right]=2\left(N_{j,\rm L}(\omega)-N_{j,\rm R}(\omega)\right)$, being $N_{j,\rm L,R}$ the boson occupation numbers for each temperature.

Therefore, in general, the total heat flux does not have a Landauer's form, but each of the terms contributing has. As we have written the total heat flux, all the terms are expressed in terms of the differences between the occupation numbers of each part and the occupation number in the right plate. This can be changed by using the identity resulting from Eq.(\ref{TotalHeatFluxEq0}) in a different way, taking as reference another of the occupation numbers.

%%%%%%%%%%%%%%%%%%%%
\section{Impact of thickness - Analytical Results}
%%%%%%%%%%%%%%%%%%%%

Once we have obtained general expressions for both the Casimir force and the heat transfer between the plates of finite width, we can recover different well-known results as limiting cases and analyze particular features to gain intuition on the physics enclosed in the general formulas.

For the case of the Casimir force, part of the features were studied in Ref.\cite{RubioLopez2017} for the case when $\beta_{\phi,\rm L}=\beta_{\phi,\rm R}\equiv\beta_{\phi}$. Now, we will summarize the relevant findings of that work and give novel generalizations of them based on the introduction of the intrinsic non-equilibrium initial state for the field.

First, the result for materials without dissipation can be recovered since ${\rm Im}(n_{i})\equiv 0$, which immediately gives $\langle\widehat{T}_{xx}^{{\rm B},\infty}\rangle^{\rm Int}_{\rm B}|_{\rm No Diss}\equiv 0$ and the Casimir force is only due to the initial conditions contribution and the regularization term. Given the intrinsic non-equilibrium state, the Casimir force in this case is given directly by the substraction of Eqs. (\ref{TxxNonEq}) and (\ref{CasimirForceICFull}), but considering real refraction indexes.

Moreover, the Lifshitz formula for the Casimir force can also be deduced from our general expressions. However, there is a subtle point that must be considered. This is how to impose Lifshitz's scenario (consisting in two half-spaces at thermal equilibrium) in our expressions. On one hand, we have to take the infinite-thickness limit as $d_{\rm L,R}\rightarrow+\infty$ and, on the other hand, we have to impose that all the temperatures are equal, $\beta_{\rm B,L}=\beta_{\rm B,R}=\beta_{\phi,\rm L}=\beta_{\phi,\rm R}\equiv\beta$. This last subtle point is crucial for derivating the correct expression for the force between half-spaces from the finite-thickness' result, since for the latter situation, three contributions enter in the expression of the force: initial conditions and baths contributions and the regularization term, each one with its own pair of temperatures. However, when taking $d_{\rm L,R}\rightarrow+\infty$, the initial conditions' term vanishes ($\langle\widehat{T}_{xx}^{{\rm IC},\infty}\rangle^{\rm Int}_{\phi}\rightarrow 0$), while the others two do not. %The fact that the initial conditions' contribution does not vanish for the infinite-thickness limit encloses the fact that the result was derived for a situation including infinite-size empty (dissipationless) regions outside the plates configuration. 
As it was shown in Ref.\cite{RubioLopez2017}, for a half-spaces configuration, there will be no initial conditions' contribution at the steady state because there is no infinite-size empty regions anywhere. In this sense, the pressure calculated and also the regularization term will be both considered with $\beta_{\rm B,L}$ and $\beta_{\rm B,R}$. For this case, having the same temperature for both half-spaces ($\beta_{\rm B,L}=\beta_{\rm B,R}\equiv\beta$) is enough to obtain Lifshitz formula, regardless on the initial state of the field. However, from a conceptual point of view, if we want to obtain Lifshitz formula as an infinite-thickness limit of the finite-width result, taking $d_{\rm L,R}\rightarrow+\infty$ together with $\beta_{\rm B,L}=\beta_{\rm B,R}\equiv\beta$ it is not enough when $\beta_{\phi,\rm L,R}\neq\beta$ in the regularization term. Clearly, by also putting $\beta_{\phi,\rm L}=\beta_{\phi,\rm R}\equiv\beta$, the total Casimir force takes the form of the Lifshitz formula.%, being the result of the sum of three contributions instead of only two (as it happens when half-spaces are analyzed from the very beginning). The limit expressions of Eqs.(\ref{CasimirForceICFull}) and (\ref{CasimirForceBathFull}) for $d_{\rm L,R}\rightarrow+\infty$ together with the Lifshitz formula (deduced in two ways, as the limit $d_{\rm L,R}\rightarrow+\infty$ and from the half-space configuration from the very beginning) can be found in Appendix \ref{LEFTF}.

Nonetheless, as we are introducing the intrinsic non-equilibrium initial state, we can go further and give also an expression for the non-equilibrium version of Lifshitz's formula, i.e., the force between two half-spaces when its temperature are different between each other. To do this, we have not only to take the limit of infinite thickness ($d_{\rm L,R}\rightarrow+\infty$) but also impose conditions over the temperatures $\beta_{\phi,\rm L,R},\beta_{\rm B,L,R}$. From the analysis done for the equilibrium case, it is clear that in the non-equilibrium case the temperatures must be grouped in left and right realizing the fact that each of the half-spaces is in local equilibrium. Therefore, we have to impose $\beta_{\phi,\rm L}=\beta_{\rm B,L}\equiv\beta_{\rm L}$ and $\beta_{\phi,\rm R}=\beta_{\rm B,R}\equiv\beta_{\rm R}$. As it is shown in Appendix \ref{LEFTF}, the infinite-thickness limit of Eq.(\ref{CasimirForceBathFull}) is given by Eq.(\ref{FBLargeD}) while Eq.(\ref{CasimirForceICFull}) vanishes. %As it is mentioned in the Appendix, it is worth noting that $F_{\rm C}^{\rm IC}[a,d\rightarrow+\infty,\beta_{\phi,\rm L},\beta_{\phi,\rm R}]=F_{\rm C}^{\rm IC}[a,d\rightarrow+\infty,\beta_{\phi,\rm L}]$ due to the fact that we are calculating the force over the left plate. If the calculation is done over the right plate, an analogous result would be obtained for this contribution, which in this case would be independent of $\beta_{\phi,\rm L}$. However, it is clear that the forces over both half-spaces have the same magnitude but opposite directions.

Then, by setting $\beta_{\phi,\rm L}=\beta_{\rm B,L}\equiv\beta_{\rm L}$ and $\beta_{\phi,\rm R}=\beta_{\rm B,R}\equiv\beta_{\rm R}$, the total Casimir force for the limit of infinite thickness ($d_{\rm L,R}\rightarrow+\infty$) in a non-equilibrium scenario is:

\begin{eqnarray}
F_{\rm C}\left[a,d_{\rm L,R}\rightarrow+\infty,\beta_{\rm L},\beta_{\rm R},\beta_{\rm L},\beta_{\rm R}\right]=\int_{0}^{+\infty}d\omega~\omega&\Bigg[&\coth\left(\frac{\beta_{\rm L}\omega}{2}\right)\left(1-\frac{\left[1-|r_{n_{\rm L}}|^{2}\right]\left[1+|r_{n_{\rm R}}|^{2}\right]}{|1-r_{n_{\rm L}}r_{n_{\rm R}}~e^{i2\omega a}|^{2}}\right)\nonumber\\
&+&\coth\left(\frac{\beta_{\rm R}\omega}{2}\right)\left(1-\frac{\left[1-|r_{n_{\rm R}}|^{2}\right]\left[1+|r_{n_{\rm L}}|^{2}\right]}{|1-r_{n_{\rm L}}r_{n_{\rm R}}~e^{i2\omega a}|^{2}}\right)\Bigg],
\label{NonEqLifshitz}
\end{eqnarray}

\noindent which is the generalization of Lifshitz's formula for the case of non-equilibrium, from which the usual Lifshitz's formula is obtained by simply setting $\beta_{\rm L}=\beta_{\rm R}\equiv\beta$.

It is worth noting that the chosen prescription to obtain the Casimir force in this non-equilibrium situation gives the correct expression, while the approach in which the force is calculated from the difference of the radiation pressures at each side of a given plate, gives an incorrect result in this scenario but a correct one in the equilibrium case.

On the other hand, for the heat transfer between the plates, similar analysis can be done exposing different conceptual properties than for the force.

A first crucial difference is that this quantity needs no regularization term since it is, from the beginning, a substraction of the radiations traveling in each directions. Moreover, as we showed in Eq.(\ref{TotalHeatFluxEq0}), the total heat flux, in equilibrium, vanishes. This is achieved since the contributions cancel between each other. Nevertheless, it should be noted that having $\beta_{\phi,\rm L}=\beta_{\phi,\rm R}=\beta_{\phi}\neq\beta_{\rm B}=\beta_{\rm B,L}=\beta_{\rm B,R}$ does not give vanishing total heat flux from the formula, which is also physically true since the scenario is an out of equilibrium one.

Regarding the contributions, it should be noted that the initial conditions' contribution $Q_{\infty}^{\rm IC}$ basically measures the asymmetry between the blackbody radiations that reach the configuration from the left and the right with different temperatures. In other words, the heat transfer associated to the initial conditions' contribution gives the difference between the radiations coming from left and right after passing the plate corresponding to the side that they come. It is the net difference between the blackbody radiations coming from the outside of the plates' configuration after interacting with the plates. In fact, that contribution for $d_{\rm L,R}=0$ recovers exactly Stefan law of blackbodies heat exchange.

On the other hand, the heat transfer associated to the baths is the difference in the radiations generated by each plate that reach the other one. In this case, for $d_{\rm L,R}=0$, the contribution automatically cancels. Moreover, as it is expected, the contribution also vanishes for dissipationless materials since ${\rm Im}(n_{i})$.

For the case of identical plates (unique material and same thickness) both contributions to the heat transfer between the plates takes the form:

\begin{eqnarray}
Q^{\rm IC}_{\infty}(a,d,d,\beta_{\phi,\rm L},\beta_{\phi,\rm R})\vert^{1-\rm Mat}%&=&\int_{0}^{+\infty}dk~k\left[\coth\left(\frac{\beta_{\phi,\rm L}k}{2}\right)-\coth\left(\frac{\beta_{\phi,\rm R}k}{2}\right)\right]\left(\left|C_{-ik}\right|^{2}-\left|D_{-ik}\right|^{2}\right)\nonumber\\
&=&\int_{0}^{+\infty}dk~k\left[N_{\phi,\rm L}(k)-N_{\phi,\rm R}(k)\right]\frac{|t|^{2}\left(1-|r|^{2}\right)}{|1-r^{2}~e^{i2ka}|^{2}},
\end{eqnarray}

\begin{eqnarray}
&&Q_{\infty}^{\rm B}(a,d,d,\beta_{\rm B,L},\beta_{\rm B,R})\vert^{1-\rm Mat}=\int_{0}^{+\infty}d\omega~\omega\left[N_{\rm B,L}(\omega)-N_{\rm B,R}(\omega)\right]
\frac{|t|^{2}(1-|r|^{2})}{|1-r^{2}~e^{i2ka}|^{2}}\frac{|n+1|^{2}}{8|n|^{2}}\\
&&~~~~~~~~~~~~~~~~~~~~~~~~~\times \Big({\rm Re}(n)(e^{2\omega{\rm Im}(n)d}-1)+{\rm Re}(n)|r_{n}|^{2}(1-e^{-2\omega{\rm Im}(n)d})+2~{\rm Im}(n)~{\rm Im}\left[r_{n}(1-e^{i2\omega{\rm Re}(n)d})\right]\Big),\nonumber
\end{eqnarray}

\noindent where both contributions have the form of a Landauer-like formula. Then, it can be easily check now that, for the case of equal temperature on both sides for each contribution ($\beta_{j,\rm L}=\beta_{j,\rm R}=\beta_{j}$), the heat transfers automatically vanish. This leads to a subtle point, associated to the fact that for the situation of two identical plates (same width and material), we can have that both contributions vanishes regardless if the temperatures of the contributions is the same. In other words, we can have that the total heat transfer vanishes although $\beta_{\phi}\neq\beta_{\rm B}$. This particular feature of the heat shows the different natures of the contributions that enters the calculations in the decomposition of the field operator acting in the total Hilbert space as a sum of operators acting in each Hilbert subspaces associated to each part of the composite system (Eq.(\ref{SteadyFieldOperator})). Moreover, we can switch-on one of the contributions independently whether the other one vanish or not, allowing to a separately study of each of the contributions. If we like to switch-on the initial conditions' contribution, it is enough to set $\beta_{\phi,\rm L}\neq\beta_{\phi,\rm R}$ while $\beta_{\rm L,B}=\beta_{\rm R,B}$. If we like the contrary, it is enough to set $\beta_{\phi,\rm L}=\beta_{\phi,\rm R}$ while $\beta_{\rm L,B}\neq\beta_{\rm R,B}$.

Finally, we can say that for the case of identical plates, the total heat transfer $Q_{\infty}$ can be given by a Landauer formula by setting $\beta_{\phi,\rm L}=\beta_{\rm L,B}=\beta_{\rm L}\neq\beta_{\phi,\rm R}=\beta_{\rm R,B}=\beta_{\rm R}$:

\begin{eqnarray}
&&Q_{\infty}(a,d,d,\beta_{\rm L},\beta_{\rm R},\beta_{\rm L},\beta_{\rm R})\vert^{1-\rm Mat}=\int_{0}^{+\infty}d\omega~\omega\left[N_{\rm L}(\omega)-N_{\rm R}(\omega)\right]
\frac{|t|^{2}(1-|r|^{2})}{|1-r^{2}~e^{i2ka}|^{2}}\\
&&\times\left[1+\frac{|n+1|^{2}}{8|n|^{2}}\Big({\rm Re}(n)(e^{2\omega{\rm Im}(n)d}-1)+{\rm Re}(n)|r_{n}|^{2}(1-e^{-2\omega{\rm Im}(n)d})+2~{\rm Im}(n)~{\rm Im}\left[r_{n}(1-e^{i2\omega{\rm Re}(n)d})\right]\Big)\right].\nonumber
\end{eqnarray}

The other case of interest also for the heat transfer is the infinite-thickness case ($d_{\rm L,R}\rightarrow+\infty$). It is straightforward that when $d_{\rm L,R}\rightarrow+\infty$, the initial conditions' contribution vanishes regardless the material of both plates is the same or not, i.e., $Q_{\infty}^{\rm IC}(a,d_{\rm L,R}\rightarrow+\infty,\beta_{\phi,{\rm L}},\beta_{\phi,{\rm R}})\equiv 0$. Although a difference on the material of each plate is allowed, being the configuration asymmetric, the infinite-size of each plate cancels the contribution of the radiation impinging from outside the configuration, giving a zero initial conditions' contribution for the heat in contrast to what happen for the case of the force.

On the other hand, the baths' contribution takes the form:

\begin{eqnarray}
Q_{\infty}^{\rm B}(a,d_{\rm L,R}\rightarrow+\infty,\beta_{\rm B,L},\beta_{\rm B,R})=\int_{0}^{+\infty}d\omega~\omega\left(N_{\rm B,L}(\omega)-N_{\rm B,R}(\omega)\right)\frac{[1-|r_{n_{\rm L}}|^{2}][1-|r_{n_{\rm R}}|^{2}]}{|1-r_{n_{\rm L}}r_{n_{\rm R}}~e^{i2\omega a}|^{2}},
\label{QBInf}
\end{eqnarray}

\noindent which is a Landauer-like formula, but different from the previous case of unique material. Again, given this formal Landauer-like expression, if we take $\beta_{\rm B,L}=\beta_{\rm B,R}=\beta$, the contribution vanishes regardless the material is the same or not.

Therefore, for the infinite-thickness case ($d\rightarrow+\infty$), we can say that the total heat transfer $Q_{\infty}$ is also given by a Landauer formula, regardless on the temperature of the radiation outside the configuration, which in this case never reachs the gap between plates.

However, regardless of these cases, the general case does not corresponds to a Landauer's formula, as we commented at the end of the previous section. Moreover, even considering the same temperature for the traveling modes ($\beta_{\phi,\rm L}=\beta_{\phi,\rm R}$), the total heat flux cannot be written as a Landauer's formula when the plates have finite width. Not even setting the temperature of the traveling modes equal to zero (and therefore, having $N_{\phi,j}\equiv 0$), the total heat flux is not given by a Landauer's formula although it depends, only,  on the baths' temperatures.

In conclusion, the total heat transfer $Q_{\infty}$ is not always given by a Landauer-like formula. Moreover, it also presents different limiting behaviours than for the force, enclosing different physical aspects. The general case of finite-width plates of different materials presents both contributions ($Q_{\infty}^{\rm IC}(a,d_{\rm L},d_{\rm R},\beta_{\phi,\rm L},\beta_{\phi,\rm R})$ and $Q_{\infty}^{\rm B}(a,d_{\rm L},d_{\rm R},\beta_{\rm B,L},\beta_{\rm B,R})$) different from zero, positioning the non-equilibrium scenario ($\beta_{\phi,\rm L}\neq\beta_{\phi,\rm R}\neq\beta_{\rm B,L}\neq\beta_{\rm B,R}$) as very richful and highly nontrivial in the interplay with the thickness role. In the next section, we investigate some aspects that can be addressed by numerical analysis.

%%%%%%%%%%%%%%%%%%%%
\section{Impact of the Thickness - Numerical Results}
%%%%%%%%%%%%%%%%%%%%

Given the exact analysis done in the previous Sectrion, it is interesting to study numerically how the width of the plates combined with the non-equilibrium features included in the general result give interesting physical aspects and let us explore the impact of the thickness in dispersion phenomena.

In this sense, including the possibility of having different temperatures but the same finite width for both plates ($d_{\rm L}=d_{\rm R}\equiv d$) there are remarkable physical effects that can be interpreted within our theoretical framework.

A crucial question that initially drives this numerical analysis and that is also of experimental interest is: given the formulas for finite width plates, which is the thickness from which the value of the total force does not differ significantly from the value of infinite-width plates ($d\rightarrow+\infty$)? In other words, for which scale of thickness $d$ the value of the total force is closer to the value for $d\rightarrow+\infty$? Moreover, from the measurement of the force, for which scale of $d$ we can say that the plates act effectively as plates of infinite thickness? From which thickness a plate of finite width can be considered practically as an infinite-width plate?

On the other hand, some questions that appear related to this analysis are: Is this scale the same 
for equilibrium or non-equilibrium scenarios? It is also the same if the quantity considered is the 
heat transfer between the plates? Is the same physics for the different contributions? Moreover, 
are there other remarkable physical effects that appears for different values of the thickness in 
non-equilibrium scenarios? Are these effects tunable in some way?

\begin{figure}
\centering
\includegraphics[width=0.75\linewidth]{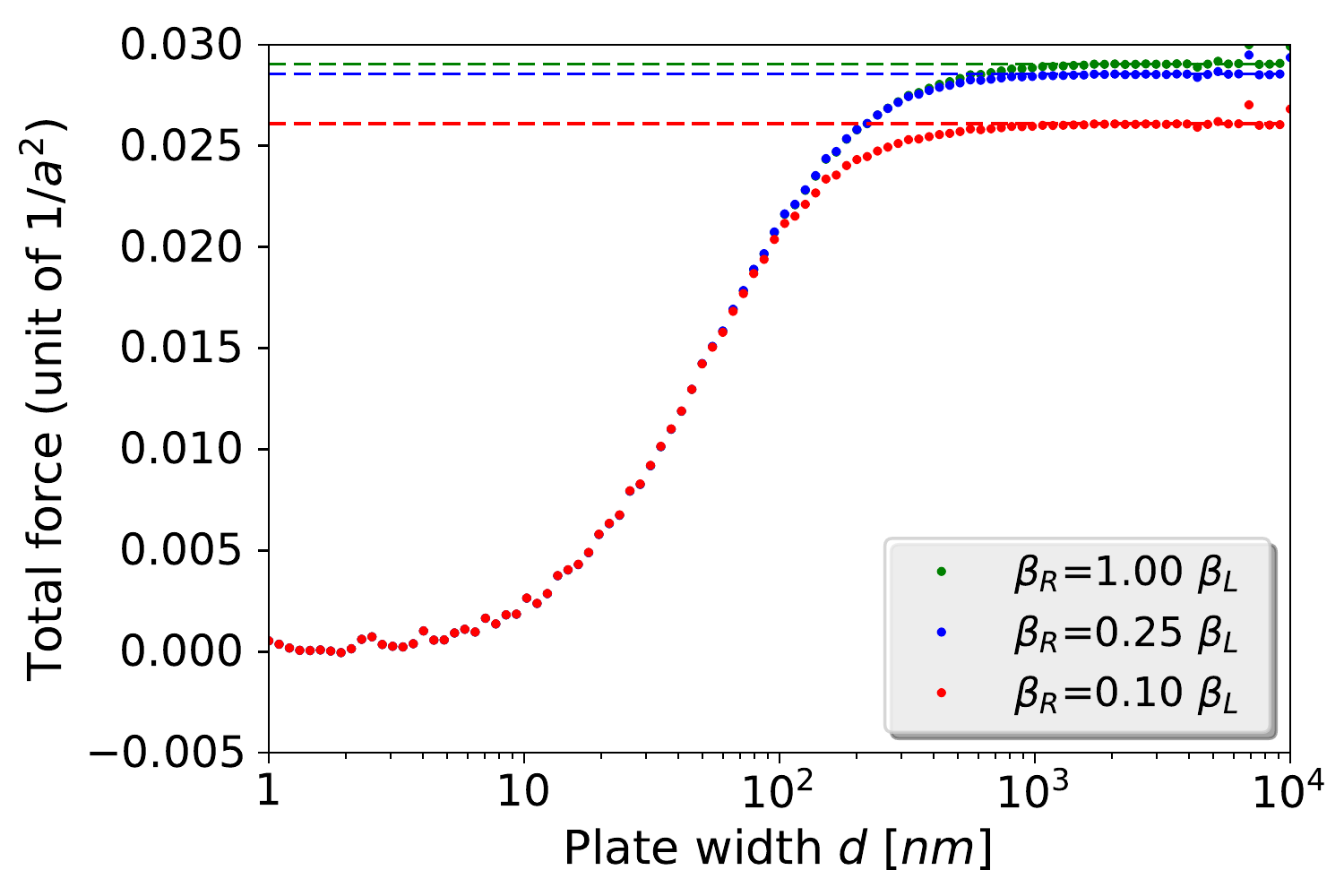}
\caption{Total Casimir force as a function of the width of the plates $d$ for a left temperature 
$T_{\rm L}=\frac{1}{\beta_{\rm L}}=300 \ \rm K$ and different right temperatures $T_{\rm R}$. 
Equilibrium and non-equilibrium cases are considered. Parameters are $\gamma_{\rm L,R}=10^{-1}/a$, 
$\omega_{0,i}=10/a$, $\omega_{\rm Pl,i}=10/a$, %$d/a=10^{2}$, 
$a=100 \ \rm nm$.}
\label{fig:1}
\end{figure}

Fig.\ref{fig:1} shows the behavior of the total Casimir force of Eq.(\ref{ForceCasimir}) as a function of the thickness $d$ for both, equilibrium and non-equilibrium scenarios. The dashed-lines  correspond to the asymptotic values ($d\rightarrow+\infty$) of the total Casimir force given by Eq.(\ref{NonEqLifshitz}). It can be observed that the scale of convergence with the thickness is of the order of the separation $a$ between the plates.

Therefore, for a given separation of the plates, a plate can be considered of infinite-width when the thickness is greater than the separation distance.

It is also worth noting that the force is maximized in the equilibrium case. Moreover, it decreases when there is more thermal difference between the plates, regardless which plate is at higher temperature. This can be physically explained since in a non-equilibrium scenario, there is a momentum exchange taking place in the region between the plates that it is not present in the equilibrium case and tends to separate the plates. Therefore, the total force between the plates decreases in its value, regardless on which plate is at a higher temperature.

\begin{figure}
\centering
\includegraphics[width=0.75\linewidth]{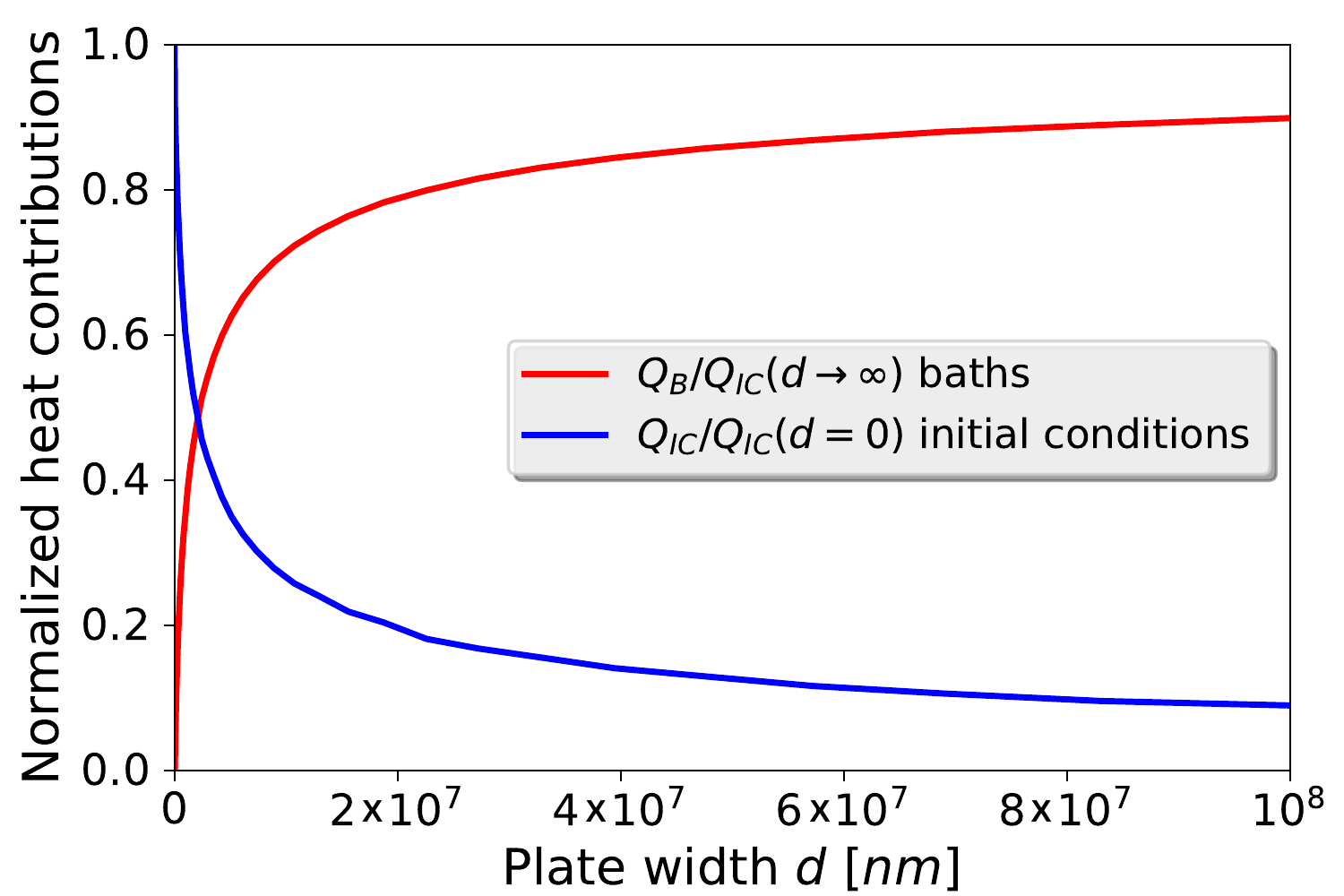}
\caption{Normalized contributions to the total heat as a function of the width of the plates for a 
right temperature $T_{\phi, {\rm R}}=T_{\rm B,R}=300 \ \rm K$ and different left temperatures 
$T_{\phi,{\rm L}}=T_{\rm B,L}$. Note that the initial conditions' contribution is normalized with 
the value at $d=0$ (which corresponds to Stefan's law), since the contributions goes to zero for 
$d\rightarrow+\infty$, while the baths' contribution is normalized in the opposite way given its 
behaviour. Parameters are $\gamma_{\rm L,R}=10^{-1}/a$, $\omega_{0,i}=10/a$, $\omega_{\rm 
Pl,i}=10/a$, %$d/a=10^{2}$, 
$a=100 \ \rm nm$.}
\label{fig:2}
\end{figure}

If we now study what happens with the heat transfer between the plates, the situation changes. In Fig.\ref{fig:2}, we observe both contributions to the total heat as a function of the plates' width $d$.

Is clear that the convergence is achieved in a very different scale than for the force. In order to 
differ in less than a $10\%$ from its asymptotic value, the thickness of the plate has to be greater 
than $10^{6}$ times the separation of the plates $a$. This means that the contributions to the heat 
transfer are more sensitive to the plates' width than the force in several orders of magnitude. 
Given the independence on the switching of each contribution, this scale could be measured by 
adjusting the physical parameters of the configuration (materials' properties and temperatures) in 
the appropriate way.

Moreover, it is worth noting that, on the one hand, for $d=0$ (corresponding to the left side of the Fig.\ref{fig:2}) we have $Q_{\infty}^{\rm B}\equiv 0$ while $Q_{\infty}^{\rm IC}\neq 0$, giving the value corresponding to the heat transfer between to distant objects at given temperatures $T_{\rm L},T_{\rm R}$, which is the one given by Stefan's law for heat exchange between two blackbodies (Eq.(\ref{BBRad})). On the other hand, for $d\rightarrow+\infty$ (corresponding to the right side of Fig.\ref{fig:2}), we have that $Q_{\infty}^{\rm IC}\equiv 0$, while $Q_{\infty}^{\rm B}$ gives Landauer-like formula expressed in Eq.(\ref{QBInf}).

\begin{figure}
\centering
\includegraphics[width=0.75\linewidth]{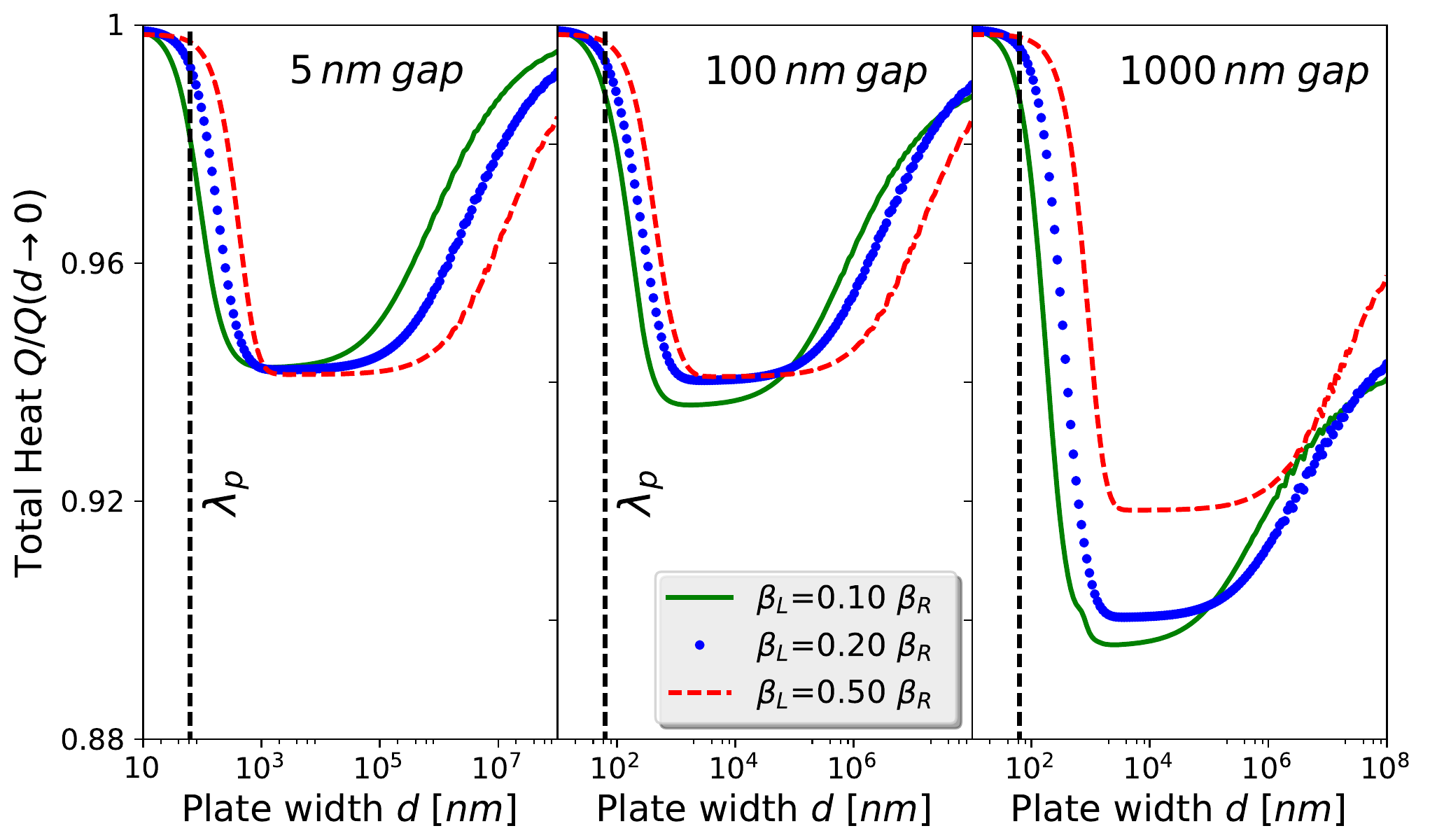}
\caption{Normalized total heat with respect to the blackbody radiation impinging ($d=0$) as a 
function of the width of the plates for a right temperature $T_{\phi, {\rm R}}=T_{\rm B,R}=300 \ K$ 
and different left temperatures $T_{\phi,{\rm L}}=T_{\rm B,L}$. Parameters are $\gamma_{\rm L,R}= 
10^{-3}{\ \rm nm}^{-1} %10^{-1}/a
$, $\omega_{0,i}=10^{-1}{\ \rm nm}^{-1} %10/a
$, $\omega_{\rm Pl,i}=10^{-1}{\ \rm nm}^{-1} %10/a
$. The dashed vertical lines corresponds to the value of the plasma wavelength $\lambda_{\rm 
Pl}\equiv\frac{2\pi c}{\omega_{\rm Pl}}\approx 63{\ \rm nm}$ valid for both plates. %$d/a=10^{2}$, 
$a=100 \ \rm nm$.
}
\label{fig:3}
\end{figure}

Considering this, we can analyze the normalized total heat flux between the plates resulting from these contributions at different separation distances $a$, obtaining Fig.\ref{fig:3} for $T_{\phi, {\rm L}}=T_{\rm B,L}\equiv T_{\rm L}$ and $T_{\phi, {\rm R}}=T_{\rm B,R}\equiv T_{\rm R}$, with $T_{\rm L}>T_{\rm R}$. The normalization is with respect to the blackbody flux corresponding to the expressions for $d=0$.

From the chosen normalization and previous comments about each contribution for $d=0$ and 
$d\rightarrow+\infty$, on one hand, we can identify the left value of each curves as the blackbody 
heat exchange between the walls of the big oven where the configuration of plates will take place, 
i.e., they correspond to $Q_{\infty}^{\rm IC}(d=0)$ for the different temperature differences and 
they are equal to 1 due to the chosen normalization. As this value is independent of the separation 
$a$, it is appropriate to take this criterion for normalizing the total heat in Fig.\ref{fig:3}. 
However, it is worth noting that for each temperature difference, the absolute values of the total 
heat even at $d=0$ are different. On the other hand, the right value of the curves correspond to 
$Q_{\infty}^{\rm B}(d\rightarrow+\infty)$. The graph then can be interpreted as the competition 
between both contributions for different values of the thickness $d$. It is worth noting that this 
competition gives rise to a minimum of the total heat transfer for a given thickness in the scale of 
the separation of the plates. Physically, the appearance of the minimum is related to the fact that 
the plates emitting radiation also act as a shield of the outside radiation coming from the walls of 
the oven. This behavior is observed when the thickness of the plates $d$ is larger than the plasma 
wavelength ($\lambda_{\rm Pl}\equiv\frac{2\pi c}{\omega_{\rm Pl}}$) for the material forming the 
plates, which in our case corresponds to $63{\rm nm}$. Then, the net result between how much 
radiation coming from the walls is screened by the plates and how much is emitted by them gives the 
total heat transfer at each thickness $d$. Thus, for small values of the thickness (with respect to 
the separation $a$), we observe that the plates screen more than they emit in the gap, giving a 
decrease in the heat flux. As the plates get thicker, the screening is increased (decreasing in the 
gap the amount of radiation coming from the walls of the oven) but also the radiation emitted by the 
plates to the gap is enhanced. For a given thickness $d$, the radiation emitted overcome the 
screening and the net heat transfer between the plates stop decreasing and begins to increase until 
the asymptotic value for $d\rightarrow+\infty$, defined only by the radiation emitted by the plates. 
The scale at which the value of the heat differs in less than $5\%$ is when the thickness is around 
$10^{7}{\rm nm}$% times the separation $a$
, but it gets longer as the separation $a$ increases.

Moreover, although the attenuation of the heat flux with respect to the infinite-thickness ($d\rightarrow+\infty$) value is of the order of $5-6\%$ in every case, with respect to the blackbody flux (when $d=0$) the percentage of attenuation varies. In fact, for a given separation, it becomes larger as soon as the temperature difference increases. At the same time, the location minimum moves to smaller orders of magnitude as the temperature difference is larger. Then, the percentage of attenuation and the location of the minimum can be tuned by increasing the temperature difference, but only in a simultaneous way.

%for the cases where $T_{\phi, {\rm L}}\neq T_{\rm B,L}$ and $T_{\phi, {\rm R}}\neq T_{\rm B,R}$ but keeping $T_{\rm i,L}>T_{\rm i,R}$. (CHEQUEAR ESTO, HACER FIGURA)

In Fig.\ref{fig:3} we also see how the total heat flux over the blacbody flux and the mentioned 
effects depend with the separation $a$. It can be seen that the attenuation could be increased by 
enlarging the separation between the plates, as well as the order of magnitude on thickness to 
achieve the attenuation becomes smaller too. This responds to the fact that for larger distances, 
the heat flux between plates provoked by the baths is lower since it involves less evanescent modes 
when increasing the separation. This is why for a distances of 5nm we observe that the attenuation 
is weaker, due to near-field enhancement of heat exchange between the plates.

\begin{figure}
\centering
\includegraphics[width=0.6\linewidth]{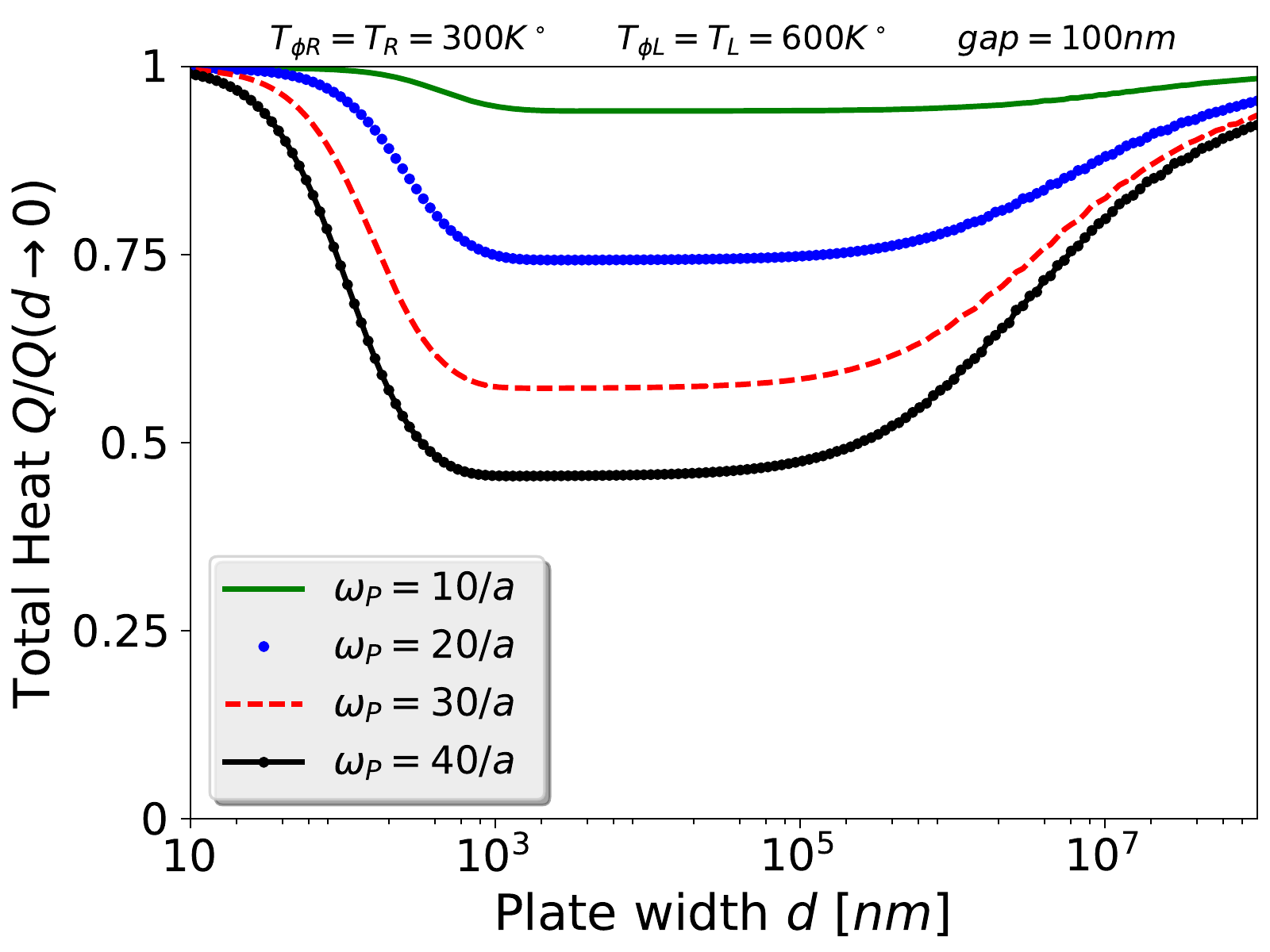}
\caption{Normalized total heat with respect to the blackbody radiation impinging ($d=0$) as a 
function of the width of the plates for for different values of plasma frequencies $\omega_{\rm 
Pl}$. The temperatures are $T_{\phi, {\rm R}}=T_{\rm B,R}=300 \ \rm K$ and $T_{\phi,{\rm L}}=T_{\rm 
B,L}=600 \ \rm K$. Parameters are $\gamma_{\rm L,R}=10^{-1}/a$, $\omega_{0,i}=10/a$, %$d/a=10^{2}$, 
$a=100 \ \rm nm$.}
\label{fig:4}
\end{figure}

Nevertheless, increasing both the difference of temperatures or the plates' separation $a$ is not efficient since it demands large thermal gradients or large separations for reaching only a $10\%$ shielding that are not desired for MEMS and NEMS devices. However, we can enhance the shielding (i.e., decrease the heat transfer) by changing the material properties. In Fig.\ref{fig:4} we can see the total normalized heat flux for a given difference of temperatures, a given separation distance and several values of the plasma frequency $\omega_{\rm Pl}$.

Increasing the plasma frequency $\omega_{\rm Pl}$ implies decreasing the plasma wavelength 
$\lambda_{\rm Pl}$, which means that the reflective properties of the plates are improved. Thus, the 
shielding of the flux related to the initial conditions' contribution is enhanced, decreasing the 
contribution more rapidly as a function of the thickness $d$, while the flux associated to the 
radiated field by each plate does not change for compensate these decays for small thicknesses. As a 
result, the minimum of the heat flux corresponds to lower percentages of the flux for $d=0$ reaching 
almost $60\%$ of attenuation when the plasma frequency is increased by four times.

In terms of materials, we can infer that for dielectrics and metals as gold this attenuation effect 
may not be significant while for metals like aluminium or platinum (that have a high energy plasma 
frequency value), this effect could not be neglected. In fact, this allows the possibility of 
measuring and tuning the effect for including it in relevant technological improvements as MEMS and 
NEMS. Indeed, as the minimum value holds approximately constant in the interval of thicknesses 
$10^3-10^5{\rm nm}$, having a $1{\rm \mu m}$ of precision on the value of the thickness of the 
plates is enough for experimentally perceive the attenuation effect when the mentioned metals are 
employed.

\begin{figure}
\centering
\includegraphics[width=0.65\linewidth]{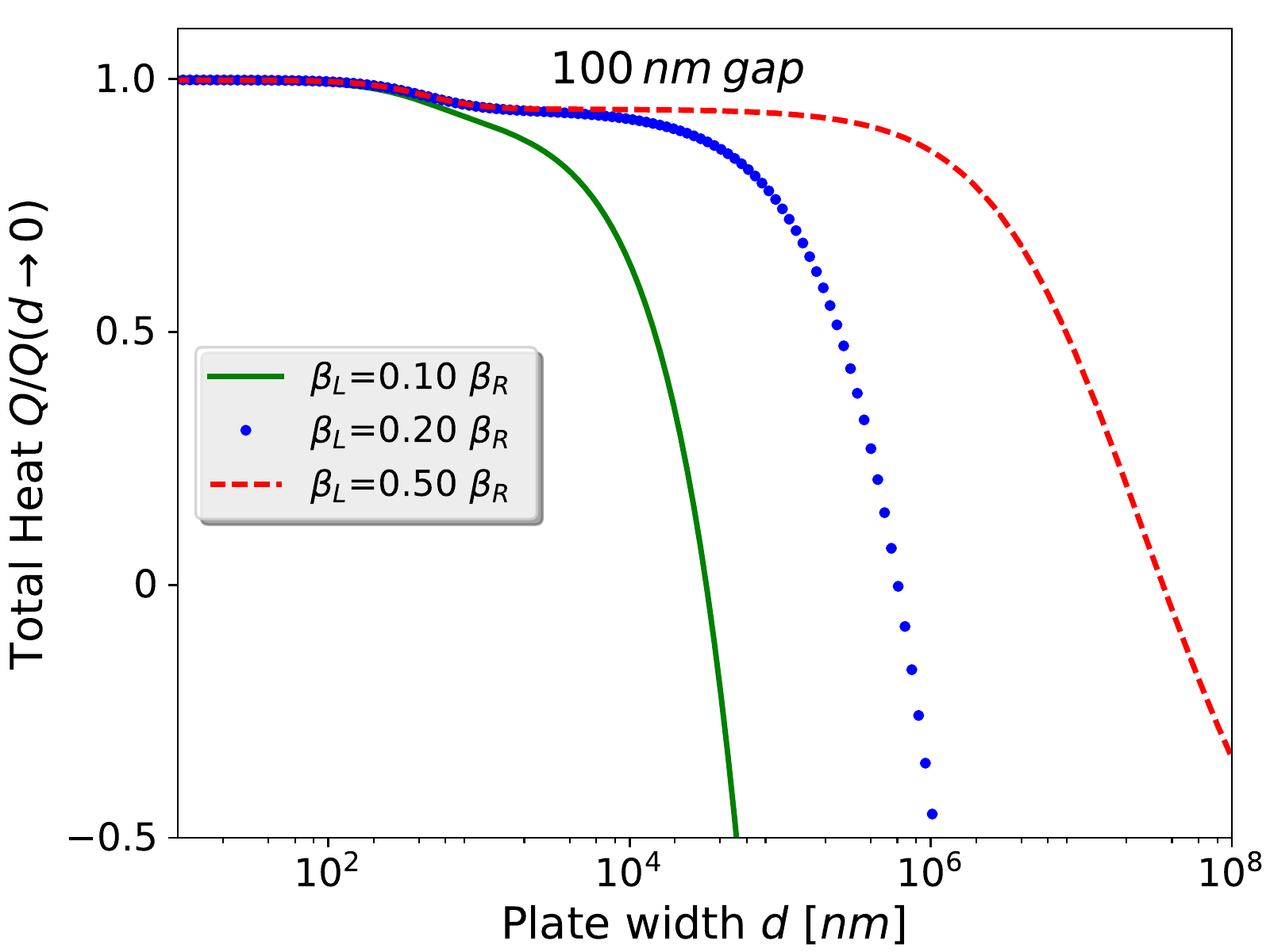}
\caption{Normalized total heat with respect to the blackbody radiation impinging ($d=0$) as a 
function of the width of the plates for crossed temperatures: $T_{\phi, {\rm L}}=T_{\rm B,R}=300 \ 
\rm K$ and different right temperatures $T_{\phi,{\rm R}}=T_{\rm B,L}$. Parameters are $\gamma_{\rm 
L,R}=10^{-1}/a$, $\omega_{0,i}=10/a$, $\omega_{\rm Pl,i}=10/a$, %$d/a=10^{2}$, 
$a=100 \ \rm nm$.}
\label{fig:5}
\end{figure}

Moreover, considering that the value of the total heat transfer at $d=0$ and $d\rightarrow+\infty$ are defined by $Q_{\infty}^{\rm IC}$ and $Q_{\infty}^{\rm B}$ respectively, it is interesting to study the situation where both quantities have opposite signs, which can be achieved by setting $T_{\phi, {\rm L}}>T_{\phi, {\rm R}}$ and $T_{\rm B,L}<T_{\rm B,R}$. For instance, by taking $T_{\phi, {\rm L}}=T_{\rm B,R}$ and $T_{\phi, {\rm R}}=T_{\rm B,L}$, we obtain Fig.\ref{fig:5}.

Considering the independence on the of the values at $d=0$ and $d\rightarrow+\infty$ and setting it 
with opposite signs, we showed that there is a thickness for which the heat transfer between the 
plates (the flux through the gap) is zero. Physically this can be understood because as a 
cancellation between the screened heat transfer due to the walls of the oven and the heat transfer 
resulting from the radiation emitted by the plates. This leads to the possibility of thermal 
shielding inside the gap where no net heat flows from one plate to the other. For the differences on 
temperature considered, the thickness at which the total heat is zero is between $10^{4}-10^{8}{\rm 
nm}$.
% times the separation between the plates. 
This shows that the value can be modified and tuned, for example, by increasing the temperature 
difference as it can be seen. It is worth noting, that varying the separation $a$ does not affect 
substantially the value of the thickness at which we obtain a zero flux.
% keeping $T_{\phi, {\rm L}}>T_{\phi, {\rm R}}$ and $T_{\rm B,L}<T_{\rm B,R}$, but with $T_{\phi, 
%{\rm L}}\neq T_{\rm B,R}$ and $T_{\phi, {\rm R}}\neq T_{\rm B,L}$.

%%%%%%%%%%%%%%%%%%%%
\section{Conclusions}
%%%%%%%%%%%%%%%%%%%%

In this work we have studied different physical aspects of dispersion phenomena in a non-equilibrium scenario, including Casimir force and heat flux. We considered a configuration of two plates of finite width $d_{\rm L,R}$ formed by materials described from a first-principles model allowing the natural introduction of dissipation, noise and temperature in the calculations. Using the formalism developed in Ref. \cite{RubioLopez2017}, we calculate the expectation value of the energy-momentum tensor operator in the steady state as a sum of two contributions, one associated to the initial conditions' of the field and the other one associated to the baths in each point of the material plates. This splitting is ensured by the fact that the baths are characterized by thermal states of temperatures $\beta_{\rm B,L},\beta_{\rm B,R}$ respectively for each plate. For the case of the field, we considered an intrinsic non-equilibrium state, where the modes traveling from left to right ($k>0$) are at a temperature $\beta_{\phi,\rm L}$, while the modes traveling from right to left ($k<0$) are at a temperature $\beta_{\phi,\rm R}$. The choice of this initial state for the field turns out to be crucial at the regularization prescription in order to obtain the non-equilibrium generalization of Lifshitz's formula (force between half-spaces) from the finite width expressions. Moreover, the correct prescription to calculate the force is the one defined from the substraction between the radiation pressure between the plates (calculated as a sum of two contributions) and the radiation pressure given without the plates in the appropriate field state, in this case, the intrinsic non-equilibrium. This is the generalization of the well-known Casimir's prescription for the case of non-equilibrium. Here, we point out that the procedure to calculate the force from the difference of the radiation pressures at each side of a given plate gives an incorrect result for the force, which is not symmetric under the exchange of the plates by doing ${\rm L}\leftrightarrow{\rm R}$. Then, we give full expressions for both, the Casimir force and the heat flux between the plates of widths $d_{\rm L},d_{\rm R}$.

For the case of the Casimir force, we reproduce particular situations as the expression for 
dissipationless materials and null widths ($d=0$). Also, we give insights to obtain the equilibrium 
Lifshitz formula (for half-spaces), but also its generalization for the non-equilibrium case (taking 
$d_{\rm L,R}\rightarrow+\infty$), which strongly depends on the intrinsic non-equilibrium state 
considered for the field. On the other hand, from the numerical analysis, we show that the scale of 
convergence in the thickness $d$ is of order of the separation $a$ in a configuration of two plates 
of same dielectric material and width. We can say that the infinite-thickness value for the force is 
effectively achieved when the width of the plates is of order of the separation between them. 
Moreover, we showed that the force decreases with the thermal imbalance, being the equilibrium value 
the maximum possible. We associate this to the fact that in a non-equilibrium scenario, there is an 
additional momentum transfer between the plates that tends to decrease the value of the net force 
between them.

For the case of the heat flux between the plates, we also give general expressions, but without requiring any regularization since the heat flux is a substraction between radiations from the very beginning. Regarding the contributions, on one hand, the initial conditions' contribution to the heat flux measure the asymmetry between the blackbody radiations reaching the configuration from each side, which reproduce Stefan's law for blackbody heat exchange when taking $d=0$. On the other hand, bath contributions are basically the difference between the radiations emitted by each plate, which we show cancels out for the case of null widths ($d=0$).

It is worth noting that in the general case, we prove that the total heat flux between the plates is not given by a Landauer-like formula. However, the expression for the total heat flux between the plates can be written in terms of differences between the boson occupation numbers at each temperature. In other words, for the general case the total heat flux has not the form of Landauer unless there are only two different values of temperature in the problem, although it can be expressed as a sum of Landauer-like terms. Moreover, even considering the vacuum state (characterized by zero temperature and number of photons) as the initial state for the field, the heat flux for finite width plates has not the Landauer's form. Nevertheless, there are some particular cases where the heat flux reduce to Landauer formulas. For example, we show that for identical plates (same material and width), both contributions result as Landauer formulas separately. If the temperature in one of the contributions agrees with the temperature in the other one, the total heat flux can be written in Landauer form (regardless on the value of the thickness). Also, we show that a Landauer formula is obtained in the case of infinite-thickness for the plates ($d_{\rm L,R}\rightarrow+\infty$), where the initial conditions' contribution goes to zero while the baths' contribution takes Landauer form even for plates of different materials.

In the numerics, for the same scenario analyzed for the force, we firstly showed that the scale of 
convergence in thickness of each contribution is several orders of magnitude greater than the case 
for the force, i.e., the thickness has to be around $10^6$ times the separation. For the total heat 
flux, on the other hand, we found two interesting behaviors as a result from the combination of both 
contributions.

For the case of considering $T_{\phi,{\rm L}}=T_{\rm B,L}=T_{\rm L}\neq T_{\rm R}=T_{\phi,{\rm 
R}}=T_{\rm B,R}$, we showed the formation of a minimum in the heat flux between the plate due to the 
opposite behaviors of each contribution. We showed that the location of this minimum can be tuned by 
varying the difference between temperatures $T_{\rm L,R}$. Moreover, the minimum implies that, for 
thicknesses of the order of the plasma wavelength of the material $\lambda_{\rm Pl}$, there is a 
shielding of the blackbody radiation impinging on the configuration, while the radiation emitted by 
the plates becomes important for large thicknesses. The relative percentage with respect to the 
asymptotic value at $d\rightarrow+\infty$ is on the order of $5-6\% $, while with respect to the 
blackbody radiation impinging ($d=0$) depends on the thermal difference and plates' separation 
considered, giving the chance to have a variable relative percentage but no significant. 
Nevertheless, taking advantage of the fact that $\lambda_{\rm Pl}$ decreases for better reflective 
materials (higher plasma frequency $\omega_{\rm Pl}$) the percentage of attenuation corresponding to 
the minimum can be tuned by changing the plasma frequency. This responds to the fact that better 
reflectivity properties result in a better shielding of the initial conditions' contribution while 
the radiation provided by the plates is not enough to compensate this effect at small thicknesses, 
increasing the attenuation to almost $60\%$ when the plasma frequency is four times the typical 
value for dielectrics. This means that a strong attenuation effect could be attainable with typical 
metals as aluminium and platinum, being of crucial importance for MEMS and NEMS devices. In other 
words, we think that our result are important in nanotechnological applications.

Considering the existence of this minimum and regardless on the material considered, we pointed out another interesting issue when considering $T_{\phi,{\rm L}}=T_{\rm B,R}\neq T_{\phi,{\rm R}}=T_{\rm B,L}$. We showed that a null heat flux can be achieved for a given thickness. This is explained by the fact that there is a cancellation of the contributions to the heat flux in the gap between the plates. The thickness for which the heat flux vanishes can be tuned by the thermal difference too. This configure a situation of thermal shielding in the gap.

As a final comment, it should be noted that these results can be easily extended to the three-dimensional scalar field, where two kinds of modes enter, the evanescent and the propagating. On the other hand, addressing the extension for the EM case could be in principle a nontrivial issue, but is in any case achievable. It is clear that the main complication will be related to the difficulties associated to quantizing the EM field, which forces us to deal with its gauge invariance and vectorial nature at a quantum framework. However, the conclusions obtained here for the scalar case will remain broadly valid for the EM field. Finally, we left as pending work the possibility of extending this analysis to include different materials and thicknesses, and also changing the temperature differences independently for each contribution.

%%%%%%%%%%%%%%%%%%%%

\acknowledgments

We would like to thank Ricardo S. Decca for useful comments and discussions on the experimental aspects that are associated to the present work. The work of AERL is supported by the Austrian Federal Ministry of Science, Research and Economy (BMWFW). The work of PMP and FCL is supported by University of Buenos Aires (UBA); CONICET, and ANPCyT.

%%%%%%%%%%%%%%%%%%%%
\appendix
%%%%%%%%%%%%%%%%%%%%

%%%%%%%%%%%%%%%%%%%%
\section{Intrinsic Non-Equilibrium Initial State of the Field}\label{INEISF}
%%%%%%%%%%%%%%%%%%%%

This appendix is devoted to comment some of the properties of the mentioned `intrinsic non-equilibrium state' for the initial state of the field. Defined by the expectation values given in Eq.(\ref{AnnihiCreaEVIntrinsicNonEq}), the state basically represents the net radiation flux given in a big oven with its vertical walls at different temperature $\beta_{\phi,\rm L}$ and $\beta_{\phi,\rm R}$ respectively.

Considering that the annihilation and creation operators for the initial conditions' contribution ($\widehat{a}_{k}(-\infty),\widehat{a}_{k}^{\dag}(-\infty)$) are the ones of the free field, we can calculate the expectation value of the Poynting vector without the presence of the plates (i.e., free space) for the intrinsic non-equilibrium state. As in this case is also valid that $\Big\langle\widehat{S}_{x}^{\text{Free}}\Big\rangle_{\phi}=-\Big\langle\widehat{T}_{x0}^{\text{Free}}\Big\rangle_{\phi}$ and having the field operator given by an expression of the form of Eq.(\ref{LongTimeFieldOperatorIC}) but with the field modes $\Phi$ replaced by plane waves $e^{\pm ikx}$ then, by using Eq.(\ref{AnnihiCreaEVIntrinsicNonEq}), we find:

\begin{eqnarray}
\Big\langle\widehat{S}_{x}^{\text{Free}}\Big\rangle_{\phi}&=&\int_{0}^{+\infty}dk~k\left[\coth\left(\frac{\beta_{\phi,\rm L}k}{2}\right)-\coth\left(\frac{\beta_{\phi,\rm R}k}{2}\right)\right]\nonumber\\
&=&2\int_{0}^{+\infty}dk~k\left[N_{\phi, \rm L}(k)-N_{\phi, \rm R}(k)\right].
\label{BBRad}
\end{eqnarray}

Both integrals are easily done as in Ref.\cite{LandsbergDeVos}, giving:

\begin{equation}
\Big\langle\widehat{S}_{x}^{\text{Free}}\Big\rangle_{\phi}=\frac{\pi^{2}}{3}\left(\frac{1}{\beta_{\phi,\rm L}^{2}}-\frac{1}{\beta_{\phi,\rm R}^{2}}\right)=\frac{\pi^{2}}{3}\left(T_{\phi,\rm L}^{2}-T_{\phi,\rm R}^{2}\right),
\end{equation}

\noindent which have the thermal dependence of the 1+1-dimensional version of Stefan's law for the heat exchange through blackbody radiation between two bodies at temperatures $T_{\phi,\rm L},T_{\phi,\rm R}$. This is the crucial point that allow us to interpret the state defined by Eq.(\ref{AnnihiCreaEVIntrinsicNonEq}) as a non-equilibrium state since it gives a heat flux even in free space. Moreover, since the radiation is blackbody-like, which is far-field radiation, we can think that all the space is inside a big oven with its walls at $x=\pm\infty$ held at different temperatures $T_{\phi,\rm L},T_{\phi,\rm R}$, causing that there is net heat transfer by radiation going from the hottest side to the other one. In other words, the intrinsic non-equilibrium state represents the state of the field when there are distant-sources in both sides emitting radiation at given different temperatures. Is clear that when $T_{\phi,\rm L}=T_{\phi,\rm R}$, the Poynting vector for free space vanishes.

On the other hand, the energy density for this state is given by:

\begin{eqnarray}
\Big\langle\widehat{T}_{00}^{\text{Free}}\Big\rangle_{\phi}=\Big\langle\widehat{T}_{xx}^{\text{Free}}\Big\rangle_{\phi}=\int_{0}^{+\infty}dk~k\left[\coth\left(\frac{\beta_{\phi,\rm L}k}{2}\right)+\coth\left(\frac{\beta_{\phi,\rm R}k}{2}\right)\right],
\label{TxxNonEq}
\end{eqnarray}

\noindent which is the typical expression for the energy density for a thermal state in free space, fully recognizable when setting $T_{\phi,\rm L}=T_{\phi,\rm R}$.

%%%%%%%%%%%%%%%%%%%%
\section{Coefficients}\label{Coeff}
%%%%%%%%%%%%%%%%%%%%

This Appendix is devoted to give the expressions of the coefficients that appears in the contributions to the Casimir force and the heat between the plates. For the given configuration of finite width plates ($d_{L,R}$), the boundary conditions on the modes were continuity of the mode and its spatial derivative at the interfaces between the material slabs and the surrounding vacuum (see Ref.\cite{RubioLopez2017} and the references therein). The coefficients then follow:

%\begin{equation}
%R_{s}^{>}=\left[r_{L}+\frac{r_{R}t_{L}^{2}~e^{-2sa}}{1-r_{L}r_{R}~e^{-2sa}}\right]e^{s(a+2d)},~~~~~T_{s}=\frac{t_{R}t_{L}~e^{2sd}}{1-r_{L}r_{R}~e^{-2sa}},
%\label{RandTCoefficients}
%\end{equation}

\begin{equation}
T_{s}=\frac{t_{R}t_{L}~e^{s(d_{\rm L}+d_{\rm R})}}{1-r_{L}r_{R}~e^{-2sa}},~~~~~C_{s}^{>}=e^{-sd_{\rm R}}~\frac{T_{s}}{t_{R}}%=\frac{t_{L}~e^{sd}}{1-r_{L}r_{R}~e^{-2sa}}
,~~~~~D_{s}^{>}=e^{-s(a+d_{\rm R})}~\frac{r_{R}}{t_{R}}~T_{s}%=\frac{t_{L}r_{R}~e^{-s(a-d)}}{1-r_{L}r_{R}~e^{-2sa}}
,
\label{CandDCoefficients}
\end{equation}

\begin{equation}
E_{s}^{>}=\frac{(n_{R}+1)}{2n_{R}}e^{s(n_{R}-1)\left(\frac{a}{2}+d_{\rm R}\right)}~T_{s},~~~~~F_{s}^{>}=\frac{(n_{R}-1)}{2n_{R}}e^{-s(n_{R}+1)\left(\frac{a}{2}+d_{\rm R}\right)}~T_{s},
\label{EandFCoefficients}
\end{equation}

\noindent where we have given the coefficients in terms of the transmission coefficients of the two plates configuration $T_{s}^{>}$. Moreover, $r_{L,R}$ and $t_{L,R}$ are the reflection and transmission coefficients for the left and right plates respectively:

\begin{equation}
r_{i}=\frac{r_{n_{i}}\left(1-e^{-2sn_{i}d_{i}}\right)}{\left(1-r_{n_{i}}^{2}e^{-2sn_{i}d_{i}}\right)},~~~~~t_{i}=\frac{4n_{i}}{(n_{i}+1)^{2}}\frac{e^{-sn_{i}d_{i}}}{\left(1-r_{n_{i}}^{2}e^{-2sn_{i}d_{i}}\right)},
\label{OnePlateRandTCoefficients}
\end{equation}

\noindent with $r_{n_{i}}=\frac{1-n_{i}}{1+n_{i}}$, the reflection coefficient of a surface of refractive index $n_{i}$.

It should be noted that the $<-$coefficients are obtained from the given ones by the interchange of $L$ and $R$ in the expressions. Considering this, it turns out that $T_{s}^{>}=T_{s}^{<}$, so that is why the superscript for this coefficient was omitted before.

%%%%%%%%%%%%%%%%%%
\section{Infinite-thickness Expressions for the Force}\label{LEFTF}
%%%%%%%%%%%%%%%%%%

This Appendix is devoted for the limit expressions that are obtained for the infinite-thickness case. This scenario will be obtained from two approaches. On one hand, by taking the limit $d_{\rm L,R}\rightarrow+\infty$ in Eqs.(\ref{CasimirForceICFull}) and (\ref{CasimirForceBathFull}), as it was similarly done in Ref.\cite{RubioLopez2017} for the case when $\beta_{\phi,\rm L}=\beta_{\phi,\rm R}\equiv\beta_{\phi}$. On the other hand, we also show that the same result can be obtained from the half-spaces scenario from the very beginning with the present formalism.

\subsection{Infinite-thickness as a limit of the finite width scenario}

To successfully take the limit of infinite width on the contributions to the total force, we are going to consider each of them separately. First of all, it is clear that the regularization term $\langle\widehat{T}_{xx}^{\rm Free}\rangle_{\phi}$ does not depend on the thickness $d$ so the term remains in the limit.

On the other hand, for the initial conditions' contribution it is enough to consider that for $d_{\rm L,R}\rightarrow+\infty$, we have that $r_{i}\rightarrow r_{n_{i}}$ while $t_{i}\rightarrow 0$. Therefore, this allow us to say that $\langle\widehat{T}_{xx}^{{\rm IC},\infty}\rangle^{\rm Int}_{\phi}\rightarrow 0$, regardless on the state considered.

For the baths' contribution, we have to take into account more subtle points when taking the limit in some combinations of factors. Considering Eq.(\ref{OnePlateRandTCoefficients}), while $t_{i}\rightarrow 0$, we also have that $|t_{i}|^{2}e^{2\omega{\rm Im}(n_{i})d_{i}}\rightarrow\frac{16|n_{i}|^{2}}{|n_{i}+1|^{4}}$. Therefore, considering the definition for the different coefficients, given in Eq.%s.(\ref{RandTCoefficients}) and 
(\ref{CandDCoefficients}), and that $1-|r_{n_{i}}|^{2}=\frac{2{\rm Re}(n_{i})}{|n_{i}+1|^{2}}$, we can write for the factor accompanying $\coth\left(\frac{\beta_{\rm B,L}\omega}{2}\right)$ in Eq.(\ref{CasimirForceBathFull}):

%\begin{equation}
%F_{\rm C}^{\rm IC}\left[a,d\rightarrow+\infty,\beta_{\phi,\rm L},\beta_{\phi,\rm R}\right]=\int_{0}^{+\infty} dk~k~\coth\left(\frac{\beta_{\phi,\rm L}k}{2}\right)\left[1+|r_{n_{\rm L}}|^{2}\right],
%\label{FICLargeD}
%\end{equation}

\begin{eqnarray}
&&\frac{{\rm Re}(n_{\rm L})}{2|t_{\rm R}|^{2}}\Big(|E_{-i\omega}^{<}|^{2}e^{-\omega{\rm Im}(n_{\rm L})a}\left[1-e^{-2\omega{\rm Im}(n_{\rm L})d_{\rm L}}\right]+|F_{-i\omega}^{<}|^{2}e^{\omega{\rm Im}(n_{\rm L})a}\left[e^{2\omega{\rm Im}(n_{\rm L})d_{\rm L}}-1\right]\nonumber\\
&&+2~\frac{{\rm Im}(n_{\rm L})}{{\rm Re}(n_{\rm L})}~{\rm Im}\Big[E_{-i\omega}^{<*}F_{-i\omega}^{<}e^{-i\omega{\rm Re}(n_{\rm L})a}\left(1-e^{-i2\omega{\rm Re}(n_{\rm L})d_{\rm L}}\right)\Big]\Big)\longrightarrow\frac{(1-|r_{n_{\rm L}}|^{2})}{|1-r_{n_{\rm L}}r_{n_{\rm R}}~e^{i2\omega a}|^{2}},
\end{eqnarray}

\noindent and the same happens for the factor accompanying $\coth\left(\frac{\beta_{\rm B,R}\omega}{2}\right)$, but interchanging L and R.

Therefore, for the contribution of the baths we have:

\begin{eqnarray}
\langle\widehat{T}_{xx}^{{\rm B},\infty}\rangle^{\rm Int}_{\rm B}\left[a,d_{\rm L,R}\rightarrow+\infty,\beta_{\rm B,L},\beta_{\rm B,R}\right]=\int_{0}^{+\infty}d\omega~\omega&\Bigg[&\coth\left(\frac{\beta_{\rm B,L}\omega}{2}\right)\frac{\left[1-|r_{n_{\rm L}}|^{2}\right]\left[1+|r_{n_{\rm R}}|^{2}\right]}{|1-r_{n_{\rm L}}r_{n_{\rm R}}~e^{i2\omega a}|^{2}}\nonumber\\
&+&\coth\left(\frac{\beta_{\rm B,R}\omega}{2}\right)\frac{\left[1-|r_{n_{\rm R}}|^{2}\right]\left[1+|r_{n_{\rm L}}|^{2}\right]}{|1-r_{n_{\rm L}}r_{n_{\rm R}}~e^{i2\omega a}|^{2}}\Bigg].
\label{FBLargeD}
\end{eqnarray}

Finally, considering this last expression and substracting it with the regularization term after setting $\beta_{\phi,\rm L}=\beta_{\rm B,L}\equiv\beta_{\rm L}$ and $\beta_{\phi,\rm R}=\beta_{\rm B,R}\equiv\beta_{\rm R}$, we obtain the force between two half-spaces at different temperature given by Eq.(\ref{NonEqLifshitz}).

Setting thermal equilibrium between the baths and the field ($\beta_{\rm B,L}=\beta_{\rm B,R}=\beta_{\phi,\rm L}=\beta_{\phi,\rm R}\equiv\beta$), the total force reads:

\begin{eqnarray}
F_{\rm C}\left[a,d_{\rm L,R}\rightarrow+\infty,\beta,\beta,\beta,\beta\right]%&=&F_{\rm C}^{\rm IC}\left[a,d\rightarrow+\infty,\beta\right]+F_{\rm C}^{\rm B}\left[a,d\rightarrow+\infty,\beta,\beta\right]\nonumber\\
%&=&4\int_{0}^{+\infty}dk~k~\coth\left(\frac{\beta k}{2}\right)\frac{\left[|r_{n_{\rm R}}|^{2}|r_{n_{\rm L}}|^{2}-{\rm Re}\left(r_{n_{\rm L}}r_{n_{\rm R}}~e^{i2\omega a}\right)\right]}{|1-r_{n_{\rm L}}r_{n_{\rm R}}~e^{i2\omega a}|^{2}}\nonumber\\
&=&-\int_{-\infty}^{+\infty}dk~k~\coth\left(\frac{\beta k}{2}\right){\rm Re}\left[\frac{r_{n_{\rm L}}(-ik)r_{n_{\rm R}}(-ik)~e^{i2ka}}{1-r_{n_{\rm L}}(-ik)r_{n_{\rm R}}(-ik)~e^{i2ka}}\right],
\end{eqnarray}

\noindent which is Lifshitz's formula.

\subsection{Infinite-thickness scenario}

The expression for the Casimir force between half-spaces at different temperatures and given distance $a$ can be also obtained by considering a half-spaces scenario from the very beginning and applying the same approach developed in Ref.\cite{RubioLopez2017}.

What it is shown there is that if in the considered scenario there are no infinite-size regions of vacuum or dissipationless material, then the steady situation is defined by the baths' contribution only (there is no initial conditions' contribution in these cases).

Therefore, for the half-spaces scenario from the very beginning, the only contribution to the energy-momentum tensor will be the baths'.

As in the finite-width case, the contribution of the baths to the field operator will be written in terms of the Green function of the given problem. Therefore, the expectation value of the components of the energy-momentum tensor can be calculated from Eq.(\ref{TMuNuB}). The information about the configuration is clearly enclosed in the Green function for the considered scenario, that can be calculated as in the method commented in Ref.\cite{RubioLopez2017}.

In the half-spaces scenario, for $-\frac{a}{2}<x<\frac{a}{2}$ and $x'<-\frac{a}{2}$, the Green function reads:

\begin{equation}
\overline{\mathfrak{G}}_{\rm Ret}(x,x',\omega)=\frac{1}{2i\omega n_{\rm L}}\left(A_{-i\omega}^{>}~e^{i\omega x}+B_{-i\omega}^{>}~e^{-i\omega x}\right)e^{-i\omega n_{\rm L}x'},
\end{equation}

\noindent where the coefficients is given by:

\begin{equation}
A_{s}^{>}=\frac{2n_{\rm L}}{(n_{\rm L}+1)}\frac{e^{s(n_{\rm L}-1)\frac{a}{2}}}{\left(1-r_{n_{\rm R}}r_{n_{\rm L}}e^{-2sa}\right)}~~~~~,~~~~~B_{s}^{>}=-\frac{2n_{\rm L}}{(n_{\rm L}+1)}r_{n_{\rm R}}\frac{e^{s(n_{\rm L}-3)\frac{a}{2}}}{\left(1-r_{n_{\rm R}}r_{n_{\rm L}}e^{-2sa}\right)}.
\end{equation}

On the other hand, for $-\frac{a}{2}<x<\frac{a}{2}$ and $\frac{a}{2}<x'$, we have the same expression for $\overline{\mathfrak{G}}_{\rm Ret}$ but exchanging L with R.

With all these considerations, it can be shown that $\langle\widehat{T}_{xx}^{{\rm B},\infty}\rangle_{\rm B}$ results equal to Eq.(\ref{FBLargeD}). Therefore, to obtain the Casimir force we have to regularize the expression by substracting the pressure without plates, which is given by Eq.(\ref{TxxNonEq}). Finally, it is clear that at the end we obtain the same expression for the non-equilibrium Casimir-Lifshitz force that we obtained from the infinite-thickness limit of the finite width scenario (Eq.(\ref{NonEqLifshitz})).

\end{document}